\documentclass[11pt]{article}

\usepackage{amsfonts,amssymb,amsthm,amsmath}
\usepackage{hyperref}
\usepackage{mathtools}
\usepackage{comment}
\usepackage[margin=1in]{geometry}
\usepackage{algorithm}
\usepackage{algpseudocode}
\usepackage{accents}

\usepackage{bm}

\usepackage{fancybox}
\usepackage{color}
\usepackage{latexsym}

\usepackage{eepic}
\usepackage{epic}
\usepackage{epsf}
\usepackage{graphics}
\usepackage{comment}
\usepackage{lineno}
\newtheorem{theorem}{Theorem}
\newtheorem*{theorem*}{Theorem}

\newtheorem{lemma}[theorem]{Lemma}
\newtheorem{conjecture}[theorem]{Conjecture}
\newtheorem{fact}[theorem]{Fact}
\newtheorem{corollary}[theorem]{Corollary}
\newtheorem{claim}[theorem]{Claim}

\newtheorem{proposition}[theorem]{Proposition}
\theoremstyle{definition}

\newtheorem{definition}[theorem]{Definition}
\newtheorem{remark}[theorem]{Remark}
\newenvironment{claimproof}[1]{\par\noindent{\em Proof of Claim.}~#1}

\newcommand{\M}{\mathbb{M}}
\renewcommand{\r}{\mathfrak{r}}

\renewcommand{\H}{\mathcal{H}}

\newcommand{\F}{\mathbb{F}}

\DeclareMathOperator{\poly}{poly}
\DeclareMathOperator{\rank}{rank}

\DeclareMathOperator{\dom}{dom}

\renewcommand{\ge}{\geqslant}
\renewcommand{\geq}{\geqslant}
\renewcommand{\le}{\leqslant}
\renewcommand{\leq}{\leqslant}

\DeclareFontFamily{OMX}{MnSymbolE}{}
\DeclareFontShape{OMX}{MnSymbolE}{m}{n}{
   <-6>  MnSymbolE5
   <6-7>  MnSymbolE6
   <7-8>  MnSymbolE7
   <8-9>  MnSymbolE8
   <9-10> MnSymbolE9
   <10-12> MnSymbolE10
   <12->   MnSymbolE12}{}
\DeclareSymbolFont{mnlargesymbols}{OMX}{MnSymbolE}{m}{n}
\SetSymbolFont{mnlargesymbols}{bold}{OMX}{MnSymbolE}{b}{n}
\DeclareMathDelimiter{\llangle}{\mathopen}{mnlargesymbols}{'164}{mnlargesymbols}{'164}
\DeclareMathDelimiter{\rrangle}{\mathclose}{mnlargesymbols}{'171}{mnlargesymbols}{'171}

\renewcommand{\angle}[1]{\langle #1 \rangle}
\newcommand{\ubar}[1]{\underaccent{\bar} #1}
\newcommand{\dangle}[1]{{\llangle} #1 {\rrangle}}

\DeclareFontEncoding{LS1}{}{}
\DeclareFontSubstitution{LS1}{stix}{m}{n}
\DeclareSymbolFont{symbols2stix}{LS1}{stixfrak} {m} {n}
\DeclareMathSymbol{\lparenless}{\mathopen} {symbols2stix}{"32}
\DeclareMathSymbol{\rparengtr}{\mathclose}{symbols2stix}{"33}

\newcommand{\newbrak}[1]{{\lparenless} #1 {\rparengtr}}

\DeclareFontEncoding{LS1}{}{}
\DeclareFontSubstitution{LS1}{stix}{m}{n}
\DeclareSymbolFont{symbols2stix}{LS1}{stixfrak} {m} {n}
\DeclareMathSymbol{\lparenless}{\mathopen} {symbols2stix}{"32}
\DeclareMathSymbol{\rparengtr}{\mathclose}{symbols2stix}{"33}

\DeclareFontFamily{OMX}{MnSymbolE}{}
\DeclareFontShape{OMX}{MnSymbolE}{m}{n}{
   <-6>  MnSymbolE5
   <6-7>  MnSymbolE6
   <7-8>  MnSymbolE7
   <8-9>  MnSymbolE8
   <9-10> MnSymbolE9
   <10-12> MnSymbolE10
   <12->   MnSymbolE12}{}
\DeclareSymbolFont{mnlargesymbols}{OMX}{MnSymbolE}{m}{n}
\SetSymbolFont{mnlargesymbols}{bold}{OMX}{MnSymbolE}{b}{n}
\DeclareMathDelimiter{\llangle}{\mathopen}{mnlargesymbols}{'164}{mnlargesymbols}{'164}
\DeclareMathDelimiter{\rrangle}{\mathclose}{mnlargesymbols}{'171}{mnlargesymbols}{'171}

\newcommand{\ey}{\mbox{\small\rm Y}}

\newcommand{\RF}{\mbox{\small\rm \textsf{RF}}}
\newcommand{\LR}{\mbox{\small\rm \textsf{LR}}}
\newcommand{\RC}{\mbox{\small\rm \textsf{RC}}} 
\newcommand{\RrSC}{\mbox{\small\rm \textsf{R-rSC}}} 
\newcommand{\IDRrSC}{\mbox{\small\rm \textsf{ID-R-rSC}}}

\newcommand{\N}{\mathbb{N}} 

\renewcommand{\angle}[1]{{\langle} #1 {\rangle}}

\newcommand{\sing}{\textsc{Singular}}

\DeclareMathSymbol{\ast}{\mathbin}{symbols}{"03}

\DeclareMathOperator{\ncrank}{\rm{ncrank}}
  
\newcommand{\tkproofpencil}{\noindent{\textbf{Proof of Theorem ~\ref{theorem-pencil-power}}}}

\newcommand{\tkproofrit}{\noindent{\textbf{Proof of Corollary ~\ref{corollary-RIT}}}}

\title{On Identity Testing and Noncommutative Rank Computation over the Free Skew Field} 
 
\author{
V. Arvind\thanks{Institute of Mathematical Sciences (HBNI), Chennai, India, \texttt{email: arvind@imsc.res.in}}  
\and Abhranil Chatterjee\thanks{Indian Institute of Technology Bombay, India, \texttt{email: abhneil@gmail.com}} 
\and Utsab Ghosal\thanks{Chennai Mathematical Institute, Chennai, India, \texttt{email: ghosal@cmi.ac.in}}
\and Partha Mukhopadhyay\thanks{Chennai Mathematical Institute, Chennai, India, \texttt{email: partham@cmi.ac.in}}
\and C. Ramya\thanks{Institute of Mathematical Sciences (HBNI), Chennai, India, \texttt{ramyac@imsc.res.in}}
}

\begin{document} 
\maketitle
\begin{abstract}
The identity testing of rational formulas (RIT) in the free skew field efficiently reduces to computing the rank of a matrix whose entries are linear polynomials in noncommuting variables \cite{HW15}. 
This rank computation problem  has  deterministic polynomial-time white-box algorithms \cite{GGOW16,IQS18} and a randomized polynomial-time algorithm in the black-box setting \cite{DM17}. 
In this paper, we propose a new approach for efficient derandomization of \emph{black-box} RIT. Additionally, we obtain results for matrix rank computation over the free skew field, and construct efficient linear pencil representations for a new class of rational expressions.  
More precisely, we show the following results: 
\begin{itemize}
\item Under the hardness assumption that the ABP (algebraic branching program) complexity of every polynomial identity for the $k\times k$ matrix algebra is $2^{\Omega(k)}$~\cite{BW05},
we obtain a subexponential-time black-box algorithm for RIT in almost general setting. This can be seen as the first ``\emph{hardness implies derandomization}" type theorem for rational formulas.  
\item 
We show that the noncommutative rank of any matrix over the free skew field whose entries have \emph{small linear pencil representations} can be computed in deterministic polynomial time. Prior to this, an efficient rank computation was only known for matrices with noncommutative \emph{formulas} as entries~\cite{GGOW20}. As special cases of our algorithm, we obtain the first deterministic polynomial-time algorithms for rank computation of matrices whose entries are noncommutative ABPs 
or \emph{rational} formulas.

\item 
Motivated by the definition given by Bergman \cite{Ber76}, we define a new class of rational functions where a rational function of inversion height at most $h$ is defined as a composition of a noncommutative r-skewed circuit (equivalently an ABP) 
with inverses of rational functions of this class of inversion height at most $h-1$ which are also disjoint. By definition, this class contains ABPs and rational formulas.
We obtain a polynomial-size linear pencil representation for this class.
As a by-product, we obtain a white-box deterministic polynomial-time identity testing algorithm for the class. 
\end{itemize}
\end{abstract}

\newpage
\section{Introduction}\label{sec:intro}
In algebraic circuit complexity the 
basic arithmetic operations are  additions, multiplications, and inverses. Using these arithmetic operations algebraic circuits compute either polynomials or rational functions.  
An important sub-area of algebraic complexity is \emph{noncommutative computation} where the multiplication of variables is not commutative  
and the set of monomials (over the variables) form a free monoid. 
If we allow only  addition and multiplication gates in the noncommutative formulas/circuits, they compute noncommutative polynomials (similar to the commutative case) in the free algebra.

In the commutative case, the role of inverses is well understood but in the noncommutative world it is quite subtle. To elaborate, it is known that \emph{any} commutative rational expression can be expressed as $fg^{-1}$ where $f$ and $g$ are two commutative polynomials~\cite{Str73}. However, noncommutative rational expressions such as $x^{-1} + y^{-1}$ or $xy^{-1}x$ cannot be represented as $f g^{-1}$ or $f^{-1} g$. If we have \emph{nested inverses} then it makes the rational expression more complicated, for example ${(z + xy^{-1}x)}^{-1}- z^{-1}$. 
Moreover, a noncommutative rational expression is not always defined on a matrix substitution. For a noncommutative rational expression $\r$, its \emph{domain of definition} is the set of matrix tuples (of any dimension) where $\r$ is defined. We denote it by $\dom(\r)$. Two rational expressions $\r_1$ and $\r_2$ are \emph{equivalent} if they agree on $\dom(\r_1)\cap \dom(\r_2)$. This induces an equivalence relation on the set of all noncommutative rational expressions (with nonempty domain of definition). It was used by Amitsur in his characterization of the \emph{universal} free skew field (denoted by $\F\newbrak{\ubar{x}}$ when the variable set is $\ubar{x}=\{x_1,x_2,\ldots,x_n\}$) and the equivalence classes are called the \emph{noncommutative rational functions} \cite{ami66}. This object plays an important role in the study of noncommutative algebra~\cite{ami66, Cohn71}, control theory~\cite{KV09}, and algebraic automata theory~\cite{vol18}. 

Computationally, rational functions are represented by noncommutative arithmetic circuits or formulas using addition, multiplication, and inverse gates \cite{HW15}. 
The \emph{inversion height} of a rational formula is the maximum number of inverse gates in a path from an input gate to the output gate. It is known that the inversion height of a rational formula of size $s$ is bounded by $O(\log s)$ \cite{HW15}. 
Hrube\v{s} and Wigderson consider the \emph{rational identity
testing} problem (RIT) of testing the equivalence of two rational formulas \cite{HW15}. It is the same as testing whether a rational formula computes the zero function in the free skew field. In other words, decide whether there exists a matrix tuple (of any dimension) such that the rational formula evaluates to nonzero on that substitution. Rational expressions exhibit peculiar properties which seem to make the RIT problem quite different from polynomial identity testing. 
The apparent lack of \emph{canonical representations} such as the sum of monomials representation for polynomials and the use of nested inverses in noncommutative rational expressions complicate the problem. For example, the rational expression $(x+xy^{-1}x)^{-1} + (x+y)^{-1} -x^{-1}$ of inversion height two is a rational identity, known as Hua's identity~\cite{Hua49}.

A second characterization of the free skew field entries was developed by Cohn~\cite{Cohn71}.
A \emph{linear pencil} $L$ of size $s$ over noncommuting variables $\ubar{x}=\{x_1, \ldots, x_n\}$ is a $s\times s$ matrix whose entries are linear forms in $\ubar{x}$ variables, i.e.\ $L = A_0 + \sum_{i=1}^n A_i x_i$, where each $A_i$ is an $s\times s$ matrix over the field $\F$. Cohn showed that for every free skew field entry $\r$ in $\F\newbrak{\ubar{x}}$, there is a linear pencil $L$ such that $\r$ is an entry of the inverse of $L$. More generally, we say that $\r$ has a \emph{linear pencil representation} of size $s$, if for vectors $\ubar{c}, \ubar{b}\in \F^s$ and $s\times s$ linear pencil $L$,
$\r =  \ubar{c}^t L^{-1} \ubar{b}$ where $\ubar{c}^t$ is the transpose of $\ubar{c}$. Hrube\v{s} and Wigderson give an efficient reduction from the RIT problem to the singularity testing problem of linear pencils~\cite{HW15}. In particular, if $\r$ is  a rational formula of size $s$, they showed that $\r$ has a linear pencil representation $L$ of size at most $2s$ such that $\r$ is defined on a matrix tuple if and only if $L$ is invertible on that tuple~\cite{HW15}. 
Using this connection, they reduce the RIT problem to the problem of testing whether a given linear pencil is invertible over the free skew field in deterministic polynomial time. 
The latter is the noncommutative $\sing$ problem, whose commutative analog is the symbolic determinant identity testing problem. The deterministic complexity of symbolic determinant identity testing is completely open in the commutative setting \cite{KI04}.  In contrast, the $\sing$ problem in noncommutative setting has deterministic polynomial-time algorithms in the white-box model due to the works of Garg et al.\ \cite{GGOW16} which is based on \emph{operator scaling} and that of  Ivanyos et al.\ \cite{IQS18} which is based on the \emph{second Wong sequence} and a constructive version of \emph{regularity lemma}. As a consequence, a deterministic polynomial-time white-box RIT algorithm follows. 

A central open problem in this area is to design an efficient \emph{deterministic} algorithm for noncommutative $\sing$ problem in the black-box case \cite{GGOW16}.
The algorithms by Garg et al.\ \cite{GGOW16} and Ivanyos et al.\ \cite{IQS18} are inherently sequential and we believe that they are unlikely to be helpful for black-box algorithm design.   
It is well-known \cite{GGOW16} that an efficient black-box algorithm (via a hitting set construction) for $\sing$ would generalize the celebrated quasi-NC algorithm for bipartite matching significantly \cite{FGT21}. There is a randomized polynomial-time black-box algorithm for this problem \cite{DM17}. 

Even for the RIT problem (which could be easier than the noncommutative $\sing$ problem), the progress towards designing efficient deterministic black-box algorithm is very limited. In fact, only very recently a deterministic quasipolynomial-time black-box algorithm for identity testing of rational formulas of \emph{inversion height two} has been designed \cite{ACM22}. It is interesting to note that in the literature of identity testing, the noncommutative $\sing$ problem and the RIT problem stand among rare examples where deterministic polynomial-time white-box algorithms are designed but for the black-box case no deterministic \emph{subexponential-time} algorithm is known.

\begin{remark}
For noncommutative polynomials computed by polynomial-size arithmetic circuits, efficient randomized polynomial identity testing algorithms are known either for polynomial degree bound or for exponential sparsity bound \cite{BW05, AJMR17}. In contrast, the complexity of testing the identity of rational circuits is completely open. In fact, even in the white-box setting we do not have a randomized subexponential-time algorithm. 
\end{remark}

\subsection{Derandomization of RIT from the hardness of polynomial identities}

In this paper, we propose a new approach to tackle the RIT problem in the black-box case under a suitable hardness assumption, which is a known conjecture in theory of Polynomial Identities (PI). This was first raised by Bogdanov and Wee \cite[Section 6.2]{BW05}. 

\begin{conjecture}\label{BW-conjecture}
The ABP complexity (i.e.\ the minimum size of an algebraic branching program) of a polynomial identity for the $k\times k$ matrix algebra $\M_k(\F)$ is $2^{\Omega(k)}$. 
\end{conjecture}

The conjecture implies that ABPs of size $s$ cannot evaluate to zero on all $O(\log s)$-dimensional matrices. 
Bogdanov and Wee \cite{BW05}  also observed that if the conjecture holds then there is an $s^{O(\log^2 s)}$-time black-box PIT for 
noncommutative ABPs\footnote{Independent of the conjecture, Forbes-Shpilka~\cite{FS13} obtained an $s^{O(\log s)}$-time black-box PIT for noncommutative ABPs.} and as supportive evidence showed that the conjecture is indeed true for \emph{normal identities} (of which the \emph{standard identity} is a special case), and the identity of \emph{algebraicity}.  

Consider the following variant of the usual hitting set definition.

\begin{definition}
For a class of rational formulas $\mathcal{R}$, we say that a hitting set $\H$ is \emph{strong} if for any formula $\r\in\mathcal{R}$, there exists a matrix tuple $\ubar{p}\in\H$ such that $\r(\ubar{p})$ is invertible. 
\end{definition}

In the following theorem, we show that an efficient derandomization of RIT is possible assuming Conjecture \ref{BW-conjecture}. This can be seen as the first ``\emph{hardness implies derandomization}" type result for rational formulas.  

\begin{theorem}\label{bw-derand-intro}
If Conjecture \ref{BW-conjecture} is true then we can construct a strong hitting set of size $(s n h(\gamma\log s)^{2h+2})^{O(h (\gamma\log s)^{2h+2})}$ for rational formulas $\r$ of size $s$ over $n$ variables and inversion height $h$ in deterministic $(s n h(\gamma\log s)^{2h+2})^{O(h (\gamma\log s)^{2h+2})}$-time for some constant $\gamma>1$. This result holds over infinite or sufficiently large finite fields and $h\leq \beta (\log s/\log \log s)$ for any $0<\beta<1$.
\end{theorem}

As a special case for $h=O(1)$, this gives a quasipolynomial-size hitting set.
To get a subexponential-size bound $2^{s^{\delta}}$ on the hitting set where $\delta$ is any constant in $(0,1)$, we 
can allow $h\leq c_{\delta}(\log s/\log \log s)$. Here $c_{\delta}\in (0,1)$ is a constant that depends on $\delta$.   

As already mentioned, the inversion height of size $s$ rational formula is bounded by $O(\log s)$~\cite{HW15}. Therefore, Theorem~\ref{bw-derand-intro} solves the RIT problem in an almost general setting.

We believe that the main interesting point about Theorem \ref{bw-derand-intro} is that it relates the black-box RIT derandomization that involves handling \emph{nested inverses} with a problem purely for noncommutative polynomials:
we can obtain a deterministic black-box RIT algorithm by showing an exponential size lower bound for ABPs computing any polynomial identity for matrix algebras. 
Over the years such hardness assumptions have proved to be useful in designing deterministic algorithms for  problems related to identity testing \cite{HS80, KI04, AMS10, DSY09, CKS18, LST21}. 

\subsubsection{Proof Sketch}

The first step in proving Theorem \ref{bw-derand-intro} is a variable reduction step that shows the identity testing of a rational formula $\r$ of inversion height $h$ can be reduced to the identity testing of another rational formula $\r'$ over $2(h+1)$ variables in a black-box manner. Notice that for noncommutative polynomials (for which $h=0$), such a reduction is standard and given by $x_i\rightarrow y_0 y^{i}_1 y_0$ where $y_0, y_1$ are new noncommutative variables. 
We prove it by induction on $h$. 
In fact, we use a stronger inductive hypothesis that roughly says that for every nonzero rational formula $\r$ of inversion height $h$, there also exists a $2(h+1)$-tuple of matrices tuple $(q_{00}, \ldots, q_{h0}, q_{01}, \ldots, q_{h1})$ such that $\r(p_1, \ldots, p_n)$ is invertible and for each $i\in [n]$, $p_i = \sum_{j=0}^h q_{j0}q^i_{j1}q_{j0}$. 
Once we assume the inductive hypothesis for inversion height $h-1$, for each rational formula $\r$ of inversion height $h$, we get a matrix tuple of the form 
$\ubar{p}=(p_1,\ldots,p_n)$ where $p_i = \sum_{j=0}^{h-1} q_{j0}q^i_{j1}q_{j0}$ such that $\r$ is defined on $\ubar{p}$. Then, we use concepts from matrix coefficient realization theory and construct the nonzero generalized series $\r(\ubar{x} + \ubar{p})$ \cite{vol18}. Now, we can use the standard bivariate encoding trick on $\r(\ubar{x}+\ubar{p})$ to complete the variable-reduction step. 

The next important step that we establish is that if Conjecture \ref{BW-conjecture} is true then for any rational formula $\r$ of size $s$ and inversion height $h$, one can find a matrix tuple $\ubar{p}$ of dimension $(\gamma\log s)^{h+1}$ (for some constant $\gamma$) such that $\r(\ubar{p})$ is an invertible matrix.  
This is done via induction on $h$ and a bootstrapping argument. For the base case, we take $h=0$. In this case the rational formula is also an ABP of size $s$ and Conjecture \ref{BW-conjecture} confirms that $\r$ is nonzero on a generic matrix tuple $\ubar{p}$ of dimension $O(\log s)$. Also $\r(\ubar{p})$ is invertible by an application of Amitsur's theorem \cite{ami66}. 
Inductively we assume that we 
can find such a matrix tuple $\ubar{q}$ of dimension $d_{h-1}\leq(\gamma\log s)^{h}$ for any rational formula $\r$ of inversion height at most $h-1$ and size at most $s$.  
An easy observation shows that given a rational formula $\r$ of inversion height $h$, $\r$ is defined on such a matrix tuple $\ubar{q}$. We again use matrix coefficient realization theory\cite{vol18} to construct the nonzero generalized series $\r(\ubar{x}+\ubar{q})$ by expanding $\r$ around the point $\ubar{q}$. Substituting the variables $x_1,x_2,\ldots,x_n$ by symbolic generic matrices over noncommuting variables $Z^{(1)}, \ldots, Z^{(n)}$ of dimension $d_{h-1}$, we observe that each entry of the output matrix $\r(\ubar{Z}+\ubar{q})$ is a recognizable series computed by a small size algebraic automaton. 

By a standard result in algebraic automata theory generally attributed to Sch\"{u}tzenberger \cite[Corollary 8.3, Page 14]{Eilenberg74},
we know that at least one of the series is nonzero even when we truncate it to a small degree. Applying the Conjecture \ref{BW-conjecture}, we infer that the truncated series is nonzero on generic matrices of dimension roughly $\approx \log (sd_{h-1})$. A simple scaling trick shows that the full (infinite)-series is also nonzero on generic matrices of same dimension. This determines the dimension of the generic matrices on which the rational formula $\r$ is nonzero. Moreover the rational formula evaluates to an invertible matrix on generic matrix substitution of that dimension. This is a consequence of Amitsur's theorem \cite{ami66}. 

Once we have these two steps, the rest of the proof is straightforward. Given nonzero $\r$ over the variables $x_1,\ldots,x_n$ of height $h$, we apply the variable reduction step to construct nonzero $\r'$ of height $h$ (and roughly of same size) over $2(h+1)$ variables $\{y_{00},y_{01}, \ldots, y_{h0}, y_{h1}\}$. Now we apply the second step that says that $\r'$ is nonzero (and hence invertible) on generic matrices over $Z$ variables of dimension $(\gamma\log s)^{h+1}$. We also make use of the fact that $\r'(\ubar{y})$ has a small-size linear pencil. To construct the final hitting set, we just need to hit two sparse polynomials 
of sparsity bound roughly $(snh(\gamma\log s)^{2h+2})^{O(h (\gamma\log s)^{2h+2})}$ and this can be done by applying the standard result of sparse polynomial hitting set construction \cite{KS01}. 

\subsection{Noncommutative rank of matrices over the free skew field}

For a matrix $M = (g_{i,j})_{m\times m}$ over the free skew field $\F\newbrak{\ubar{x}}$, its noncommutative rank(denoted by $\ncrank(M)$) is the least positive integer $r\leq m$ such that $M=PQ$ for an $m\times r$ matrix $P$ and an $r\times m$ matrix $Q$ over $\F\newbrak{\ubar{x}}$. This is also called the {\em inner rank}. If $r = m$, then $M$ is invertible in $\F\newbrak{\ubar{x}}$. 

Indeed, a fundamental result of Cohn \cite{Cohn95} showed that 
for any matrix $M = (g_{i,j})_{m\times m}$ over the noncommutative ring $\F\angle{\ubar{x}}$ such that $\ncrank(M) = r$, there exists an $m\times r$ matrix $P$ and an $r\times m$ matrix $Q$ over $\F\angle{\ubar{x}}$.

As already mentioned, the problem of computing the noncommutative rank of a linear matrix admits deterministic polynomial-time white-box algorithms~\cite{GGOW16, IQS18}. 
If the matrix entries consist of some higher degree terms, one can use Higman's trick~\cite{Hig40} to reduce it to computing rank of a linear matrix. 
Consider the following well-known example of a $2\times 2$ matrix \cite{GGOW20}:
\[
\begin{bmatrix}
1 &x\\
y &z+xy
\end{bmatrix}.
\]
Higman's trick reduces it to another $3\times 3$ linear matrix preserving the complement of the noncommutative rank in the  following way:

\[
\begin{bmatrix}
1 &x\\
y &z+xy
\end{bmatrix}
\mapsto
\begin{bmatrix}
1 &x &0\\
y &z+xy &0\\
0 &0 &1
\end{bmatrix}
\mapsto
\begin{bmatrix}
1 &x &0\\
y &z &x\\
0 &-y &1
\end{bmatrix}.
\]

However, it would not be efficient in general. In ~\cite[Proposition~A.2]{GGOW20}, the authors showed an effective use of Higman's trick to efficiently reduce it to the rank computation of a linear matrix when the entries are computed by noncommutative \emph{formulas}. 

In this paper, we address the matrix rank computation over the free skew field in a very general setting.   
In particular, we obtain an efficient reduction to the rank computation of a linear matrix even when the entries are free skew field elements computed by small linear pencils. More precisely, we show the following.

\begin{theorem}\label{thm-nc-rank}
Let $M = (g_{i,j})_{m\times m}$ be a matrix such that for each $i,j\in [m]$, $g_{i,j}$ in $\F\newbrak{x_1,\ldots, x_n}$ has a linear pencil of size at most $s$. Then, the noncommutative rank of $M$ can be computed in deterministic $\poly(m,n,s)$ time. Moreover, in deterministic $\poly(m,n,s)$ time, we can output a matrix tuple $\ubar{T}=(T_1, \ldots, T_n)$ of dimension $d$ such that the matrix the rank of matrix $M(\ubar{T})$ is $d\cdot \ncrank(M)$.
The field $\F$ could be infinite or sufficiently large finite field.   
\end{theorem}

As any noncommutative formula has a small linear pencil, our result subsumes a particular result of Garg et al.\ \cite{GGOW20} which shows the efficient matrix rank computation when the entries are noncommutative formulas. If the entries of the matrices are computed by  noncommutative ABPs, by a direct application of the algorithm due to Garg et al.\ \cite{GGOW20} we can compute the rank in deterministic quasipolynomial time as any ABP has a quasipolynomial-size formula. However, since a noncommutative ABP has a polynomial-size linear pencil~\cite{HW15}, as a direct corollary of Theorem~\ref{thm-nc-rank}, we obtain a deterministic $\poly(m,n,s)$-time algorithm for the ABP case. Moreover, since noncommutative rational formulas also have polynomial-size linear pencils~\cite{HW15}, we obtain a deterministic $\poly(m,n,s)$-time algorithm even if each entry of the matrix is computed by a rational formula.


 \subsubsection{Proof Sketch}
 
The basic principle of our proof is to reduce the problem to the rank computation of a linear matrix. However, there is no clear notion of \emph{degree reduction} for arbitrary elements over the free skew field. This forces us to find a new approach of constructing this linear matrix efficiently that can also handle a matrix of skew field entries as input. The main idea of the proof is to show that the linear pencil representation enjoys the following closure property. Let $A$ be an $m\times m$ generic matrix over $m^2$ indeterminates and let substituting each indeterminate of $A$ by a free skew field entry that also has a linear pencil of size at most $s$, we obtain $M$. We show that we can find a small linear matrix $L$ efficiently such that $\ncrank(L) = m^2s + \ncrank(M)$. Somewhat surprisingly, the construction of $L$ turned out to be relatively simple and elegant.  

There are many equivalent notions of noncommutative rank for linear matrices (for example, see \cite{IQS18, GGOW16}). A notion of particular interest is the blow-up definition that is crucial in the algorithm of Ivanyos et al.\ \cite{IQS18}. The blow-up notion enables to find a matrix tuple on which the maximum rank is achieved.  
We extend this notion and introduce a blow-up definition for noncommutative rank (denoted by $\ncrank^*$) of matrices with free skew field entries. We show that for any matrix $M$ of free skew field entries, $\ncrank(M) = \ncrank^*(M)$. Introduction of the blow-up definition allows us to find efficiently the matrix tuple $\ubar{T}$ of dimension $d$ such that the rank of $M(\ubar{T})$ is $d\cdot\ncrank(M)$. One can view the blow-up definition in this case as an extension of the theory developed by Derksen and Makam \cite{DM17} for the linear case. This extension could be of independent mathematical interest.

\subsection{Linear pencil representations for a new class of rational functions}


The study of linear pencils seem to be the key in understanding several basic questions in rational function theory \cite{HW15, GGOW20, IQS18, vol18, DM17}. In this section, our main motivation is to understand the relation between the linear pencil representations of rational functions and the representations using basic arithmetic operations. 
Let $\RF, \LR, \RC$ be the class of polynomial-size rational formulas, the class of rational functions that have polynomial-size linear pencil representations, and the class of polynomial-size rational circuits. Hrube\v{s} and Wigderson \cite{HW15} prove an exponential size lower bound on the size of the rational formulas computing an entry of the inverse of a symbolic matrix.  Moreover, they show that each entry of the inverse of a symbolic matrix is computable by a rational circuit of polynomial size. Therefore, the current known relation is $\RF\subset \LR \subseteq \RC$.




Following Bergman \cite{Ber76}, a noncommutative rational function $\r(\ubar{x})$ of inversion height at most $h$ can be inductively defined as $\r(\ubar{x}) = f(x_1, \ldots, x_n, g^{-1}_1, \ldots, g^{-1}_m)$, where $f$ is a noncommutative polynomial and $g_1, \ldots, g_m\in\F\newbrak{\ubar{x}}$ are rational functions of inversion height $\leq h-1$. 
Using this, we give the following definition.

\begin{definition}\label{defn-0}
A \emph{rational r-skewed circuit} of inversion height $0$ is a noncommutative r-skewed circuit\footnote{Usually in the literature they are called right-skew circuits. For the purpose of this paper, we prefer referring to them as \emph{right-skewed} and reserve the word ``skew" for the \emph{skew field}.} which is also a noncommutative ABP. Inductively, we define $\r(\ubar{x}) = f(x_1, \ldots, x_n, g^{-1}_1, \ldots, g^{-1}_m)$ as a rational r-skewed circuit of inversion height at most $h$ if $f(\ubar{x}, y_1, \ldots, y_m)$ is a noncommutative r-skewed circuit ($m\geq 0$) and for each $i\in [m]$, $g_i(\ubar{x})$ is a rational r-skewed circuit of inversion height $\leq h-1$. 
\end{definition}

Let $\RrSC$ be the class of all rational functions computable by polynomial-size rational r-skewed circuits. Inspecting 
the polynomial size rational circuit for symbolic matrix inverse \cite{HW15}, one can notice that each entry of the inverse of a polynomial-size symbolic matrix can indeed be computed by a polynomial-size rational r-skewed circuit. Hence $\LR \subseteq \RrSC$. What is the exact expressive power of the class $\LR$? In particular, is it true that $\LR=\RrSC$? It now suffices to show that $\RrSC \subseteq \LR$. While we are unable to answer this completely, we exhibit such a containment under additional structural restriction.

\begin{definition}\label{defn-1}
An \emph{inversely disjoint rational r-skewed circuit} of inversion height $0$ is a noncommutative r-skewed circuit (which is also an ABP). Inductively, we define $\r(\ubar{x}) = f(x_1, \ldots, x_n, g^{-1}_1, \ldots, g^{-1}_m)$ as an inversely disjoint rational r-skewed circuit of inversion height at most $h$ if $f(\ubar{x}, y_1, \ldots, y_m)$ is a noncommutative r-skewed circuit ($m\geq 0$) and for each $i\in [m]$, $g_i(\ubar{x})$ is a inversely disjoint rational r-skewed circuit of inversion height $\leq h-1$ and for all $i\ne j$, the circuits of $g_i$ and $g_j$ are disjoint. 
\end{definition}

Let $\IDRrSC$ be the class of rational functions computed by polynomial-size \emph{inversely disjoint r-skewed circuits}.  This class contains rational formulas, ABPs. We are able to give polynomial-size linear pencil representations for this class. 

\begin{theorem}\label{theorem-pencil-power}
Over any field, an inversely disjoint rational r-skewed circuit of size $s$ has a linear pencil representation of size 
$O(s^2)$ which can be computed in deterministic polynomial time from the given circuit. 
\end{theorem}

This gives the following containment: 
\[
\RF\subseteq \IDRrSC\subseteq \LR\subseteq \RrSC\subseteq \RC,
\]
where we know at least one of the first two containment is proper.
We do not know any unconditional separation between $\RF$ and $\IDRrSC$. 
This question is somewhat similar in spirit to the separation of noncommutative formulas and ABPs which is still open \cite{Ni91, Chatterjee21, LST22}. However, a simple inductive argument shows that a function of inversion height $h$ in $\IDRrSC$ can be computed by 
rational formula of size $s^{O(h\log s)}$. By the standard argument, a noncommutative r-skewed circuit of size $s$ can be computed by a formula of size $s^{O(\log s)}$. 
Consider an inversely disjoint r-skewed circuit $\r(\ubar{x}, g_1^{-1}, \ldots, g_m^{-1})$ where each $g_i\in\IDRrSC$ of inversion height $\leq h-1$ for each $1\leq i\leq m$. 
Inductively, each $g_i$ has a rational formula of size $s^{O((h-1)\log s)}$. 
Therefore, 
the size of the rational formula computing $\r$ can be at most $s^{O(h \log s)}$. If $h = O(\log s)$, we then have a quasipolynomial-size formula simulation for this class. However, unlike rational formulas~\cite{HW15}, it is not clear whether $h$ can be taken as $O(\log s)$ for a general inversely disjoint r-skewed circuit of size $s$.

Using Theorem~\ref{theorem-pencil-power}, the following 
corollary is obtained by the application of rank computation algorithm in \cite{IQS18}. For the black-box case, we can apply the algorithm in~\cite{DM17}. In the proof of the corollary we also mention how to apply the algorithm in \cite{IQS18} for the black-box case and get an efficient randomized algorithm over the finite fields also. 

\begin{corollary}\label{corollary-RIT}
Let $\F$ be infinite or any sufficiently large field. For an inversely disjoint rational r-skewed circuit of size at most $s$ and over $n$ variables, we can decide whether it computes zero in $\F\newbrak{\ubar{x}}$ or not in deterministic $\poly(s,n)$ time in white-box, and in randomized $\poly(s,n)$ time in black-box. 
\end{corollary} 

\subsubsection{Proof Sketch}
As the key component, the proof uses a composition lemma that computes an efficient linear pencil for $f(\ubar{x}, g_1^{-1}, \ldots, g_m^{-1})$ from the linear pencils of $f(\ubar{x}, \ubar{y})$ and $g_1^{-1}, \ldots, g_m^{-1}$. It turns out that the proof of this composition result is more subtle than the usual proofs of the linear pencil constructions for rational formulas \cite{HW15, vol18}. 

We first elaborate on the composition lemma. 
Let $L$ be an $s\times s$ linear pencil over $x_1, \ldots, x_n$ and
$y_1, \ldots, y_m$. Let $f_{i, j} = (L^{-1})_{i, j}$ for $i,j \in
[s]$. Let $g_1, \ldots, g_m$ be rational
functions over $x_1, \ldots, x_n$ such that each $g_k$ has a linear pencil $L_k$ of size at most
$s'$. Then we can construct a single linear pencil $\widetilde{L}$ of
size at most $ms' + m + 2s^2+ s$ in $\poly(s', s, m, n)$-time such that
\[
(\widetilde{L}^{-1})_{2s^2 + \widehat{s}+i, 2s^2 + \widehat{s}+j} = f_{i,j} (\ubar{x}, g^{-1}_1, \ldots, g^{-1}_m)
\quad\text{ for } i,j\in[s], \text{ where } \widehat{s} = ms' + m.
\]
Given a rational function $\r$ computed by an inversely disjoint rational r-skewed circuit of size at most $s$, we consider the rational function $\r^{-1}$ which is still in the same class (with inversion height increased by one). Using the composition result, we construct a linear pencil of size $O(s^2)$ for $\r^{-1}$. Notice that $\r(\ubar{x},\ubar{y})$ is a polynomial computed by an ABP or a r-skewed circuit and it has a polynomial-size linear pencil~\cite{HW15}. Using a standard idea, $\r^{-1}$ also has a small linear pencil $L$ which we use as the input to the composition lemma along with the inductively constructed linear pencils for $g_1, \ldots, g_m$. 

The final linear pencil $\widetilde{L}$ which is the outcome of the composition lemma has the additional property that for any matrix tuple $\r(\ubar{p})$, $\r^{-1}(\ubar{p})$ is defined if and only if $\widetilde{L}(\ubar{p})$ is invertible. Since $\r\neq 0$ if and only if $\r^{-1}$ is defined~\cite{ami66}, we can now use the algorithm for noncommutative $\sing$ problem \cite{IQS18} on the linear pencil $\widetilde{L}$ to check the identity of $\r$.    

\subsection*{Organization} 
In Section~\ref{section-preli}, we mainly provide  brief background on linear pencils and its connection with the rational identity testing problem, and also present some results in matrix coefficient realization theory. 
We prove Theorem \ref{bw-derand-intro} in Section \ref{sec:cond-derandom}. 
The proof of Theorem \ref{thm-nc-rank} is given in Section \ref{sec:rank-computation}. 
We give the proof of Theorem~\ref{theorem-pencil-power} 
in Section~\ref{expressive-power}. We state some open questions in Section \ref{sec:open}.

\section{Preliminaries}
\label{section-preli}

\subsection{Linear pencils and rational functions}

Let $\F$ be a field. A linear pencil $L$ of size $s$ over noncommuting $\ubar{x}=\{x_1,\ldots,x_n\}$ variables is a $s\times s$ matrix where each entry is a linear form in $\ubar{x}$. That is,  $L = A_0 + \sum_{i=1}^n A_i x_i$ where each $A_i$ in $\M_s(\F)$. Evaluation of a linear pencil at a matrix tuple $\ubar{p} = (p_1, \ldots, p_n)$ in $\M^n_m(\F)$ is defined using the Kronecker (tensor) product: $L$ evaluated at $\ubar{p}$ is $A_0\otimes I_m + \sum_{i=1}^n A_i\otimes p_i$. 

Given a linear pencil $L$, the noncommutative $\sing$ problem is to decide whether there is a tuple $\ubar{p}$ in $\M^n_m(\F)$ of $m\times m$ matrices for some $m$ such that the output matrix $L$ evaluated at $\ubar{p}$ is invertible. 

A rational function $\r$ in $\F\newbrak{\ubar{x}}$ has a linear pencil representation $ L$ of size $s$ if $\r = \ubar{c}^t L^{-1}\ubar{b}$ 
for vectors $\ubar{c},\ubar{b}\in \F^s$. 
Following is the re-statement of Proposition 7.1 proved in \cite{HW15}.  

\begin{proposition}\label{prop:pencil-representation}
Let $\r$ be a rational function given by a rational formula of size $s$. Then $\r$ can be represented $(L^{-1})_{i,j}$ for $i,j\in [s]$ where $L$ is a linear pencil of size at most $2s$. Moreover, $\r$ is nonzero if and only if $L$ is invertible.  
\end{proposition}

Clearly in the above proposition the choice for $\ubar{c},\ubar{b}$ are the indicator vectors $e_i$ and $e_j$. 

We also use the following classical result of Amitsur \cite{ami66} in this paper. 

\begin{theorem}[\cite{ami66}]\label{amitsur-domain}
Let $\r$ be a rational function which is nonzero on $\M_k(\F)$ where $\F$ is infinite or any sufficiently large field. Then $\r(\ey_1,\ldots,\ey_n)$ is an invertible matrix in $\M_k(\F(\ubar{\ey}))$ where 
$\ey_1, \ldots, \ey_n$ are generic indeterminate matrices of dimension $k$. 
\end{theorem}

\begin{remark}\label{rmk:amitsur-anyfield}
Usually Theorem \ref{amitsur-domain} is stated over infinite fields. However it can be adapted over any sufficiently large finite field $\F$ using the techniques in~\cite{IQS18}. We briefly discuss it here. For details we refer the reader to ALGORITHM~1 in \cite{IQS18}. Define the field $\F'$ by adjoining a $k^{th}$ root $\zeta$ to $\F$ i.e.\ $\F'=\F[\zeta]$. Then construct a $\F'(Z_1, Z)$ basis $\Gamma=\{C_1, \ldots, C_{k^2}\}$ of $\M_k(\F'(Z_1, Z))$ such that $\F'(Z_1, Z^k)$-linear span of $\Gamma$ is a central division algebra over $\F'(Z_1, Z^k)$. Here $Z_1, Z$ are two independent formal variables. Using that we can see that $\r$ is invertible on a generic linear combination of $\Gamma$. Now by a standard argument the generic variables can be fixed from $\F$ (assuming that $\F$ is sufficiently large) to obtain a matrix tuple $\ubar{T}$ such that $\r(\ubar{T})$ is invertible. This also implies that $\r(\ubar{Y})$ is invertible where $\ubar{Y}$ is a generic matrix tuple of dimension $k$.     
\end{remark}

\subsection{Algebraic branching programs (ABPs)}
\begin{definition}\label{defn:abp}
An algebraic
  branching program (ABP) is a layered directed acyclic graph with
one in-degree-$0$ vertex called \emph{source}, and one out-degree-$0$
vertex called \emph{sink}. Its vertex set is partitioned into layers
$0, 1, \ldots, d$, with directed edges only between adjacent layers
($i$ to $i+1$). The source and the sink are in layers zero and $d$,
respectively. Each edge is labeled by a linear form over $\F$
in variables $\{x_1,\ldots,x_n\}$. The polynomial computed by
the ABP is the sum over all source-to-sink directed paths of the
product of linear forms that label the edges of the path. The maximum
number of nodes in any layer is called the width of the algebraic
branching program. The size of the branching program is taken to be
the total number of nodes.
\end{definition}

Equivalently, an ABP of width $w$ and $d$ many layers can be defined as an entry of a product of $d$ many linear matrices of size at most $w$. Therefore, the polynomial $f$ computed by an ABP is of form $(M_1\cdots M_d)_{i,j}$ for some $i,j\in [w]$.

\begin{proposition}\label{prop:abp-pencil}
An ABP of size $s$ has a linear pencil of size at most $2s$ from the following construction:  
\[
L_f = 
\begin{bmatrix}
I_w &-M_1\\
& I_w & -M_2\\
&&\ddots &\ddots\\
&&&I_w &-M_d\\
&&&&I_w
\end{bmatrix}.
\]
The ABP is computed in the upper right corner. 
\end{proposition}

This construction is well-known and also used in \cite{HW15}. 







\subsection{Matrix Inverse}\label{block-matrix}

Let $P$ be a $2\times 2$ block matrix shown below. 
\begin{equation*}
\quad P = 
\begin{bmatrix}
p_1  &p_2 \\
p_3 &p_4	
\end{bmatrix}
\end{equation*}
where $p_1$ is invertible and $p_2$ and $p_3$ can be any rectangular matrices and $(p_4 - p_3p^{-1}_1p_2)$ is also invertible. Then we note that the inverse of $P$ has the following structure \cite{HW15}.

\begin{equation}\label{2by2inverse}
\quad P^{-1} = 
\begin{bmatrix}
p_1^{-1} (I + p_2 (p_4 - p_3p^{-1}_1p_2)^{-1} p_3 p_1^{-1})   & -p_1^{-1} p_2 (p_4 - p_3p^{-1}_1p_2)^{-1} \\
-(p_4 - p_3p^{-1}_1p_2)^{-1} p_3 p_1^{-1} &(p_4 - p_3p^{-1}_1p_2)^{-1}	
\end{bmatrix}
\end{equation}

If $p_3 = 0$, then $P^{-1}$ has a simpler structure.
\begin{equation}\label{2by2-special}
P^{-1} = \begin{bmatrix}
p^{-1}_1 &-p_1^{-1}p_2p^{-1}_4\\
0 &p^{-1}_4
\end{bmatrix}.
\end{equation}

Hrube\v{s} and Wigderson use Equation~\ref{2by2inverse} to compute each entry of the matrix inverse recursively by a small rational circuit.
\begin{theorem}\cite[Theorem~2.4]{HW15}
Each entry of an $s\times s$ symbolic matrix is computable by a rational circuit of size $O(s^{\omega})$ where $\omega$ is the exponent of matrix multiplication.
\end{theorem} 

\begin{remark}\label{remark:matrix-inverse-by-r-skew}
 We observe that the same construction also yields a polynomial-size rational r-skewed circuit as defined in Definition~\ref{defn-0} for the matrix inverse.
Inspecting Equation~\ref{2by2inverse}, we just need to compute the entries of $p^{-1}_1$ and $(p_4-p_3p^{-1}_1p_2)^{-1}$ and after that the remaining computation is straightforward. Notice that, in the composition step while replacing each $y_i$ by $g^{-1}_i$, Definition~\ref{defn-0} allows any $g_i$ to be a sub-circuit of some $g_j$. Therefore, we can reuse the r-skewed circuit computing each entry of $p^{-1}_1$ and follow the same recursive construction to obtain a rational r-skewed circuit of size $O(s^{\omega})$. 
\end{remark}
\subsection{Recognizable series}
A comprehensive treatment is in the book by Berstel and Reutenauer~\cite{BR11}. We will require the following concepts. Recall that $\F\dangle{\ubar{x}}$ is the formal power series ring over a field $\F$. A series $S$ in $\F\dangle{\ubar{x}}$ is \emph{recognizable} if it has the following linear representation: for some integer $s$, there exists a row vector $\ubar{c}\in \F^{1\times s}$, a column vector $\ubar{b}\in \F^{s\times 1}$ and an $s\times s$ matrix $M$ whose entries are homogeneous linear forms over $x_1, \ldots, x_n$ i.e. $\sum_{i=1}^n \alpha_ix_i$ such that $S = \ubar{c}^t\left(\sum_{k\geq 0}M^k\right)\ubar{b}$. Equivalently, $S = \ubar{c}^t(I-M)^{-1}\ubar{b}$. We say, $S$ has a representation $(\ubar{c}, M, \ubar{b})$ of size $s$ \footnote{In the language of weighted automata, the matrix $M$ is the transition matrix for the series $S$.}.

The following theorem is a basic result in algebraic automata theory.
\begin{theorem}\label{thm:sch-finite}
A recognizable series with representation $(\ubar{c}, M, \ubar{b})$ of size $s$ is nonzero if and only if $\ubar{c}^t\left(\sum_{k\leq s-1}M^k\right)\ubar{b}$ is nonzero.
\end{theorem}

It has a simple linear algebraic proof~\cite[Corollary 8.3, Page 145 ]{Eilenberg74}. This result is generally attributed to Sch\"{u}tzenberger. For the purpose of this paper, the theorem is used to apply that the truncated series is computable by a small noncommutative ABP therefore reducing zero-testing of recognizable series to the identity testing of noncommutative ABPs.  
  

\subsection{Matrix coefficient realization theory} 
The noncommutative rational functions
lack a canonical form. If a noncommutative rational function is analytic (or defined) at a matrix point, then (matrix coefficient)-realization theory offers a representation of the noncommutative rational function around that point. This is also common in automata theory and control theory.  
For a detailed exposition of this theory, see the work of Vol\v{c}i\v{c}~\cite{vol18}. 

Recall that, $\M_m(\F)$ is the $m\times m$ matrix algebra over $\F$. A \emph{generalized word} or a \emph{generalized monomial} in $x_1,\ldots, x_n$ over $\M_m(\F)$ allows the matrices to interleave between variables. More formally, a generalized word over $\M_m(\F)$ is of the following form: $a_0 x_{k_1}a_2\cdots a_{d-1}x_{k_d}a_{d}$ where $a_i\in \M_m(\F)$. A generalized polynomial over $\M_m(\F)$ is obtained by a finite sum of generalized monomials in the ring $\M_m(\F)\angle{\ubar{x}}$. Similarly, a generalized series over $\M_m(\F)$ is obtained by infinite sum of generalized monomials in the ring $\M_m(\F)\dangle{\ubar{x}}$.

A generalized series (resp. polynomial) $S$ over $\M_m(\F)$ admits the following canonical description. Let $E=\{e_{i,j}, 1\leq i,j\leq m\}$ be the set of matrix units. Express each coefficient matrix $a$ in $S$ in the $E$ basis by a $\F$-linear combination and then expand $S$. Naturally each monomial of degree-$d$ in the expansion looks like $e_{i_0,j_0} x_{k_1} e_{i_1,j_1} x_{k_2} \cdots e_{i_{d-1},j_{d-1}} x_{k_d} e_{i_d,j_d}$ where $e_{i_l,j_l}\in E$ and $x_{k_{l}}\in \ubar{x}$. We say the series $S$ (resp. polynomial) is identically zero if and only if it is zero under such expansion i.e. the coefficient associated with each generalized monomial in the canonical representation is zero.   

The evaluation of a generalized series over $\M_m(\F)$ is defined on any $k'm\times k'm$ matrix algebra for some integer $k'\geq 1$ \cite{vol18}. To match the dimension of the coefficient matrices with the matrix substitution, we use an inclusion map $\iota: \M_m(\F)\to \M_{k'm}(\F)$, for example, $\iota$ can be defined as $\iota(a) = a\otimes I_{k'}$ or $\iota(a) = I_{k'}\otimes a$. We now define the evaluation of a generalized series (resp. polynomial) over $\M_m(\F)$ in the following way. Any degree-$d$ generalized word  $a_0x_{k_1}a_1\cdots a_{d-1}x_{k_d}a_{d}$ over $\M_m(\F)$ on a matrix substitution $(p_1,\ldots, p_n)\in \M^n_{k'm}(\F)$ evaluates to $$ \iota(a_0) p_{k_1} \iota(a_1)\cdots \iota(a_{d-1}) p_{k_d} \iota(a_d) $$under some inclusion map $\iota:\M_m(\F)\to \M_{k'm}(\F)$.  In ring theory, all such inclusions are known to be compatible by the Skolem-Noether theorem~\cite[Theorem 3.1.2]{row80}. Therefore, if a series $S$ is zero with respect to some inclusion map $\iota: \M_m(\F)\to \M_{k'm}(\F)$, then it must be zero w.r.t. any such inclusions. The equivalence of the two notions of zeroness follows from the proof of \cite[Proposition 3.13]{vol18}. 

We now recall the definition of a recognizable generalized series from the same paper.
\begin{definition}
A generalized series $S$ in $\M_m(\F)\dangle{\ubar{x}}$ is \emph{recognizable} if it has the following linear representation. For some integer $s$, there exists a row-tuple of matrices $\boldsymbol{c}\in (\M_m(\F))^{1\times s}$, and $\boldsymbol{b}\in (\M_m(\F))^{s\times 1}$ and an $s\times s$ matrix $M$ whose entries are homogeneous generalized linear forms over $x_1, \ldots, x_n$ i.e. $\sum_{i=1}^n p_ix_iq_i$ where each $p_i,q_i\in \M_m(\F)$ such that 
$S = \boldsymbol{c}(I-M)^{-1}\boldsymbol{b}$. We say, $S$ has a linear representation $(\boldsymbol{c}, M, \boldsymbol{b})$ of size $s$ over $\M_m(\F)$.
\end{definition} 

In \cite{vol18}, Vol\v{c}i\v{c} shows the following result.

\begin{theorem}~\cite[Corollary~5.1, Proposition 3.13]{vol18}\label{rit-connection}
Given a noncommutative rational formula $\r$ of size $s$ over $x_1, \ldots, x_n$ and a matrix tuple $\ubar{p}\in \M^n_m(\F)$ in the domain of definition  of $\r$, $\r(\ubar{x} + \ubar{p})$ is a recognizable generalized series with a representation of size at most $2s$ over $\M_m(\F)$. Additionally, $\r(\ubar{x})$ is zero in the free skew field if and only if $\r(\ubar{x}+\ubar{p})$ is zero as a generalized series. 
\end{theorem}

\begin{proof}
For the first part, see Corollary 5.1 and Remark 5.2 of \cite{vol18}. 

To see the second part, let $\r(\ubar{x})$ is zero in the free skew field. Then the fact that $\r(\ubar{x} + \ubar{p})$ is a zero series follows from Proposition 3.13 of \cite{vol18}. 
If $\r(\ubar x)$ is nonzero in the free skew field, then there exists a matrix tuple $(q_1, \ldots, q_n)\in \M^n_l(\F)$ such that $\r(\ubar q)$ is nonzero. W.l.o.g. we can assume $l = k'm$ for some integer $k'$. Fix an inclusion map $\iota: \M_m(\F)\to \M_{k'm}(\F)$. Define a matrix tuple $(q'_1, \ldots, q'_n)\in \M^n_{k'm}(\F)$ such that $q'_i = q_i - \iota(p_i)$. Therefore, the series $\r(\ubar x +\ubar p)$ on $(q'_1, \ldots, q'_n)$ evaluates to $\r(\ubar{q})$ under the inclusion map $\iota$, hence nonzero~\cite[Remark~5.2]{vol18}. Therefore, $\r(\ubar x + \ubar p)$ is also nonzero.
\end{proof}
\begin{remark}\label{remark:explicit-point}
More explicitly we can say the following which is already outlined in \cite[Section 5]{vol18}. For  inclusion map $\iota :\M_m(\F)\to \M_{k'm}(\F)$ 
\[
\r(\ubar{q} + \iota(\ubar{p})) = \iota(\boldsymbol{c}) \left(I_{2sk'm} - \sum_{j=1}^n \iota(A^{x_j})(\ubar{q})\right)^{-1}\iota(\boldsymbol{b}).
\]
\end{remark}

We also note down a few basic facts. The following is easy to show and also noted in \cite{vol18}. 

\begin{fact}\label{fact:psi}
Let $\r(\ubar{x}+\ubar{p})$ be a generalized series where $\ubar{p}$ consists of matrices in $\M_m(\F)$. If we replace each $x_i$ by a generic matrix over noncommuting variables $(y^{i}_{j,k})_{1\leq j,k \leq m}$, then we get a nonzero matrix over the $\ubar{y}$ variables. More precisely, the map $\psi(x_i) = (y^{i}_{j,k})_{1\leq j,k \leq m}$ is identity preserving. 
\end{fact}

Another easy fact is the following. 
\begin{fact}\label{fact:recognizable}
Let $\r(\ubar{x}+\ubar{p})$ has a linear representation $\boldsymbol{c}(I-M)^{-1}\boldsymbol{b}$ of size $s$. Then each entry of $\r(\psi(\ubar{x}) + \ubar{p})$ is a recognizable series with transition matrix $M(\psi(x_1), \ldots, \psi(x_n))$ of size $sm$. More precisely, the $(i,j)^{th}$ entry of $\r(\psi(\ubar{x}) + \ubar{p})$  has a representation $(\ubar{c}_i, M(\psi(\ubar{x})), \ubar{b}_j)$ where $\ubar{c}_i$ and 
$\ubar{b}_j$ are the $i^{th}$ row and $j^{th}$ column of $\boldsymbol{c}$ and $\boldsymbol{b}$ repectively. 
\end{fact}

\section{Derandomization of RIT from the Hardness of Polynomial Identities}\label{sec:cond-derandom}

In this section, we present a new approach to derandomize (almost general) RIT efficiently in the black-box setting and prove Theorem \ref{bw-derand-intro}.
Given a noncommutative polynomial $P(x_1, \ldots, x_n) \in\F\angle{x_1, \ldots, x_n}$, there is a well-known trick to reduce the identity testing of $P$ to the identity testing of a bivariate polynomial $P'(y_0,y_1)$ over the noncommuting variables $y_0, y_1$ by the substitution $x_i\leftarrow y_0 y^i_1 y_0$ for $1\leq i\leq n$. 

For a rational formula $\r(\ubar{x})$, such a variable reduction step preserving identity is not immediate. Our first result in this section reduces the identity testing of an $n$-variate rational formula of inversion height $h$ to the identity testing of a rational formula of inversion height $h$ over $2(h+1)$ variables. But before that we record a simple fact. 

\begin{fact}\label{fact:point-of-defn}
Given any rational formula $\r'$ of of inversion height at most $h-1$ and size at most $s$, if we can find a matrix tuple such that $\r'$ is invertible on that matrix tuple, then for a rational formula $\r$ of size at most $s$ and inversion height $h$, we can find a matrix tuple where $\r$ is defined.   
\end{fact}

\begin{proof}
Let $\mathcal{F}$ be the collection of all those inverse gates in the formula $\r$ such that for every $g \in \mathcal{F}$, the path from
the root to $g$ does not contain any inverse gate. For each $g_i \in \mathcal{F}$, let $h_i$ be the sub-formula input to
$g_i$. 
Consider the formula $\r' = h_1 h_2 \cdots h_k$ (where $k = |\mathcal{F}|$) which is of size at most $s$ since for each
$i$ and $j$, $h_i$ and $h_j$ are disjoint. Clearly, $\r'$ is of inversion height at most $h-1$. So if we find a point $\ubar{q}$ such that $\r'(\ubar{q})$ is invertible then $\r$ is defined at that point $\ubar{q}$.   
\end{proof}

Now we state and prove the variable reduction lemma for rational formulas. 

\begin{lemma}\label{lem:variable-reduction}
Let $\r(x_1, \ldots, x_n)$ be a rational formula of inversion height $h$. Then, there exists a $2(h+1)$ variate rational formula $\r'$ of inversion height $h$ over the variables $\{y_{j0}, y_{j1} : 0\leq j\leq h\}$ such that $\r$ is zero in $\F\newbrak{\ubar{x}}$ if and only if $\r'$ is zero in $\F\newbrak{\ubar{y}}$. Moreover, $\r'$ is obtained from $\r$ by substituting $x_i$ by $\sum_{j=0}^h y_{j0}y^i_{j1}y_{j0}$ for $1\leq i\leq n$. 
\end{lemma}

\begin{proof}
The proof is by induction on the inversion height $h$. 
In fact we use a stronger inductive hypothesis:
For every nonzero rational formula of inversion height $h$, there also exists a matrix tuple $(p_1, \ldots, p_n)$ and a collection of matrices $\{q_{00}, \ldots, q_{h0}, q_{01}, \ldots, q_{h1}\}$ such that $\r(p_1, \ldots, p_n)$ is invertible and for each $i\in [n]$, $p_i = \sum_{j=0}^h q_{j0}q^i_{j1}q_{j0}$.

It is true for noncommutative polynomials for which $h=0$. It is already mentioned that the substitution $x_i\leftarrow y_{0} y^i_{1} y_{0}$ reduces the identity testing of $P(\ubar{x})$ to the identity testing of $P'(y_0, y_1)$. Moreover, by Theorem \ref{amitsur-domain}, we know  that we can find matrices $q_0, q_1$ such that the bivariate polynomial $P'(q_0, q_1)$ evaluates to an invertible matrix. Since $P(q_0 q_1 q_0, q_0 q^2_1 q_0, \ldots, q_0 q^n_1 q_0)=P'(q_0, q_1)$, we establish the base case of the induction.  

Inductively, suppose that it is true for any formula of inversion height $h-1$. Now consider a nonzero rational formula $\r(x_1, \ldots, x_n)$ of inversion height $h$. From the inductive hypothesis and 
Fact~\ref{fact:point-of-defn}, there exists a matrix tuple $(p_1, \ldots, p_n)$ and a collection of matrices $\{\tilde{q}_{00}, \ldots,\tilde{q}_{(h-1)0}, \tilde{q}_{01}, \ldots, \tilde{q}_{(h-1)1}\}$ such that $\r(p_1, \ldots, p_n)$ is defined and for each $i\in [n]$, $p_i = \sum_{j=0}^{h-1} \tilde{q}_{j0}\tilde{q}^i_{j1}\tilde{q}_{j0}$.  Let the dimension of each $p_i$ be $m$. Therefore, $\r(\ubar{x}+\ubar{p})$ is also a nonzero generalized series by Theorem \ref{rit-connection}. Replacing each $x_i$ by $y_0y^i_1y_0$, we obtain a nonzero bivariate generalized series and suppose it is nonzero for $y_0 = \widehat{q_{h0}}$ and $y_1 = \widehat{q_{h1}}$ of some dimension $km$ for an integer $k$.
Notice from Section~\ref{section-preli} that a generalized series is zero if and only if the coefficient of every monomial in the canonical representation is zero. Therefore the bivariate substitution $x_i \to y_0y^i_1y_0$ preserves the nonzeroness of a generalized series.
Therefore,
\[
\r(\widehat{q_{h0}}\widehat{q_{h1}}\widehat{q_{h0}} + \iota(p_1), \ldots,  {\widehat{q_{h0}}}{\widehat{q_{h1}}}^n\widehat{q_{h0}} + \iota(p_n) )
\]
is also nonzero. Notice that, $\iota(p_i) = \sum_{j=0}^{h-1} \iota(\tilde{q}_{j0})(\iota(\tilde{q}_{j1}))^i\iota(\tilde{q}_{j0})$  for the inclusion map $\iota$ from $\M_m(\F)\rightarrow \M_{km}(\F)$.
We can now define $\r'$ substituting each $x_i$ in $\r$ by $\sum_{j=0}^h y_{j0}y^i_{j1}y_{j0}$. Clearly, $\r'$ is nonzero. By Theorem \ref{amitsur-domain}, $\r'$ is also invertible for some matrix tuple $\ubar{q}$ of same dimension. Hence $\r(p_1,\ldots,p_n)$ is invertible for $p_i = \sum_{j=0}^h q_{j0}q^i_{j1}q_{j0}$. 
\end{proof}

Next we show that if Conjecture \ref{BW-conjecture} is true then any rational formula of size $s$ and inversion height $h\leq \beta (\log s/\log\log s)$ for $\beta\in(0,1)$, is nonzero on a matrix tuple of dimension $(\gamma\log s)^{h+1}$ for some constant $\gamma$. 
\vspace{0.5cm}

\begin{lemma}\label{lem:dimension-bound}
Let $\r(x_1, \ldots, x_n)$ be a nonzero rational formula of size $s$ and inversion height $h \leq \beta (\log s/\log\log s)$ for any constant $0<\beta<1$. Then, Conjecture \ref{BW-conjecture} implies that there is a matrix tuple $(p_1, \ldots, p_n) \in \M^n_m(\F)$ such that $\r(p_1, \ldots, p_n)$ is invertible and $m = (\gamma \log s)^{h+1}$ for some constant $\gamma>1$. 
\end{lemma}

\begin{proof}
~The proof is by induction on $h$. For the base case $h=0$, Conjecture~\ref{BW-conjecture} implies that the noncommutative formula is nonzero on generic $c\log s$ (for some constant $c$) dimensional matrix tuple $(Z_1, \ldots, Z_n)$ where $Z_i= (z^{(i)}_{\ell, k})_{1\leq \ell, k\leq c\log s}$. Also Theorem~\ref{amitsur-domain} says that the formula evaluates to an invertible matrix $M(Z)$ on substituting $x_i$ by $Z_i$. Now using standard idea, random substitution to the variables in $Z_1,\ldots, Z_n$ yields such a matrix tuple. 

Inductively assume that we have already proved the dimension bound on the witness of the invertible image for rational formulas of inversion height at most $h-1$. Let the dimension of the matrices be $d_{h-1}$.  Now given a rational formula $\r$ of size $s$ and inversion height $h$, observe that $\r$ is defined on some $d_{h-1}\times d_{h-1}$ matrix tuple $\ubar{q}$ using Fact \ref{fact:point-of-defn}.  

Then by Theorem \ref{rit-connection}, $\r(\ubar{x} + \ubar{q})$ can be represented by a recognizable generalized series of size at most $2s$ such that $\r(\ubar{x})$ is nonzero if and only if $\r(\ubar{x} + \ubar{q})$ is nonzero. Using Fact \ref{fact:psi}, apply the $\psi$ map on the variables such that $\psi(x_i)$ substitutes the variable $x_i$ by a matrix of fresh noncommuting variables $z^{(i)}_{j,k}$ for $1\leq j,k\leq d_{h-1}$. 

Using Fact \ref{fact:recognizable}, observe that we get a matrix of recognizable series and each such recognizable series can be represented by an automaton of size at most 
$\hat{s}\leq 2 s d_{h-1}$. Since $\psi$ preserves identity, one of such recognizable series will be nonzero. So w.lo.g, let the series be $S_{1,1}$ computed at $(1,1)^{th}$ entry is nonzero. Let the transition matrix for $S_{1,1}$ is $M_{1,1}$.
Then using Theorem \ref{thm:sch-finite}, the truncated finite series  $\tilde{S}_{1,1}=\ubar{c}^t\left( \sum_{k\leq \hat{s}-1} M_{1,1}^k \right) \ubar{b}$ is nonzero, which is a noncommutative ABP.

If Conjecture \ref{BW-conjecture} is true then $\tilde{S}_{1,1}$ will be nonvanishing on a matrix tuple $\ubar{p}$ of dimension $O(\log \hat{s})$. Now by the following simple scaling trick, we show that the infinite series
$S_{1,1}$ is nonzero at a matrix tuple of dimension $c\log \hat{s}$. 

\begin{claim}\label{claim:scaling-trick}
We can find a matrix tuple $\ubar{p}'$ which is a scalar multiple of $\ubar{p}$ such that 
$S_{1,1}(\ubar{p}')$ is nonzero. 
\end{claim}

\begin{proof}
Let $\tau$ be a commutative variable and consider the matrix tuple, \[\tau\ubar{p}=(\tau p^{\{1\}}_{1,1},\ldots, \tau p^{\{1\}}_{d_{h-1}, d_{h-1}}, \ldots, \tau p^{\{n\}}_{1,1}, \ldots, \tau p^{\{n\}}_{d_{h-1}, d_{h-1}}).\]  Observe that $M_{1,1}(\tau\ubar{p}) = \tau M_{1,1}(\ubar{p})$. From the definition of the series $S_{1,1}$,
\[
S_{1,1}(\ubar{z}) = \tilde{S}_{1,1}(\ubar{z}) + \sum_{i\geq \hat{s}} \ubar{c}^t M_{1,1}^{i} \ubar{b}.
\]
Let $d$ be the dimension of the matrices in the tuple $\ubar{p}$. 
We now evaluate $S_{1,1}$ at $\tau\ubar{p}$ to get the following: 
\[
S_{1,1}(\tau\ubar{p}) = \tilde{S}_{1,1}(\tau\ubar{p})+ \sum_{i\geq \hat{s}}\tau^i  \cdot \left((\ubar{c}\otimes I_d)^t\cdot  M_{1,1}^i(\ubar{p})\cdot (\ubar{b}\otimes I_d)\right). \]
Since $\tilde{S}_{1,1}(\ubar{p})\neq 0$,
we have that  $S_{1,1}(\tau\ubar{p})$ evaluates to a nonzero matrix  whose entries are power series in the variable $\tau$.

It is also true that $S_{1,1}(\tau\ubar{p}) = (\ubar{c}\otimes I_d)^{t} \cdot (I- M_{1,1}(\tau\ubar{p}))^{-1}\cdot (\ubar{b}\otimes I_d)$ which is rational expression in $\tau$ where the degrees of the numerator and denominator polynomials are bounded by $\poly(\hat{s}, d)$. Hence we need to avoid only $\poly(\hat{s}, d)$ values for $\tau$ such that   
$S_{1,1}(\tau\ubar{p})$ is defined and nonzero. 
\end{proof} 

The above argument shows that for a specific value $\tau_0$ for the parameter $\tau$, the generalized series $\r(\ubar{x} + \ubar{q})$ evaluates to nonzero on a matrix tuple $(N_1(\tau_0) + \iota(q_1), \ldots, N_n(\tau_0) + \iota(q_n))$ where $N_i$ is obtained from the matrix $(z^{(i)}_{j,k})_{1\leq j,k\leq d_{h-1}}$ by substituting the variables $(z^{(i)}_{j,k})_{1\leq j,k\leq d_{h-1}}$ by $\tau_0 p^{(i)}_{j,k}$. Also $\iota$ is the inclusion map  $\iota : \M_{d_{h-1}}(\F)\rightarrow \M_{d d_{h-1}}(\F)$ defined as $\iota({q_i}) = {q_i}\otimes I_{d}$. 

Hence $\r$ is nonzero on generic matrix tuples of dimension $d_h=d d_{h-1} \leq c d_{h-1} \log(s d_{h-1})$.
Inductively assume that $d_{h-1}\leq (2c\log s)^{h}$. Since $h\leq \beta (\log s/\log\log s)$, we can observe that $s\geq d_{h-1}$. Using this we get that $d_h \leq c (2c\log s)^{h}\log(s^2)$ and that yields $d_h\leq (2c \log s)^{h+1}$. We take $\gamma=2c$. 

Therefore by Theorem~\ref{amitsur-domain} $\r(\ubar{x})$ evaluates to an invertible matrix on substituting $x_i$ by generic matrices of dimension $(\gamma\log s)^{h+1}$. 
\end{proof}



Now we are ready to show that if Conjecture \ref{BW-conjecture} is true, then we can find a subexponential-size hitting set for rational formulas of size $s$ and inversion height  up to $c'(\log s/\log\log s)$ for a suitable constant $c'$ that depends on the exponent of the subexponential function.  

\vspace{0.5cm}
\noindent\textbf{Proof of Theorem~\ref{bw-derand-intro}.}
Let $\r(x_1, \ldots, x_n)$ be a rational formula of inversion height $h$ and size $s$. Consider, $\r'(y_{00}, y_{01}, \ldots, y_{h0}, y_{h1})$ obtained from $\r$ by substituting $x_i$ by $\sum_{j=0}^h y_{j0}y^i_{j1}y_{j0}$ for $1\leq i\leq n$. From Lemma~\ref{lem:variable-reduction}, we know that $\r(\ubar{x})$ is nonzero if and only if $\r'(\ubar{y})$ is nonzero. Moreover, $\r'$ has a rational formula of size at most $s'$ which is of $O(s n h)$. Therefore, $\r'$ must be invertible on $d_h\times d_h$ generic matrix substitution where $d_h \leq (\gamma\log s')^{h+1}$ from Lemma~\ref{lem:dimension-bound}. Using Proposition \ref{prop:pencil-representation}, we know that $\r'$ has a linear pencil $L'$ of size at most $2s'$. W.l.o.g, assume that $\r'$ is computed at the $(1,1)^{th}$ entry of $L'^{-1}$. 

Hence, if we substitute the variables $y_{00}, y_{01}, \ldots, y_{0h}, y_{1h}$ by $d_h\times d_h$ generic matrices $\{Z^{(i,0)}, Z^{(i,1)} : 0\leq i\leq h)\}$ (over commuting variables), the $(1,1)^{th}$ block of $L'^{-1}(\ubar{Z})$ will be of form $\frac{M'(\ubar{Z})}{\det(L'(\ubar{Z}))}$ where $\det(L'(\ubar{Z}))$ is a polynomial of degree at most $2s' (\gamma\log s')^{h+1}$. Further, each entry of the matrix $M'$ is a cofactor of $L'(\ubar{Z})$ and therefore it is a polynomial over the $\ubar{Z}$ variables of degree at most $2s' (\gamma\log s')^{h+1}$. This shows that $\det(M'(\ubar{Z}))$ is a nonzero polynomial of degree at most $2s' (\gamma\log s')^{2h+2}$. 

The sparsity of $\det(L'(\ubar{Z}))$ and $\det(M'(\ubar{Z}))$ are bounded by $\kappa=(s'(\gamma\log s)^{2h+2})^{O(h (\gamma\log s)^{2h+2})}$. Now we can use standard sparse polynomial hitting set for $\kappa$-sparse polynomials to hit both the polynomials \cite{KS01}. This gives us a strong hitting set $\H'$ for $\r'$.

Consequently, we get a strong hitting set of same size for $\r$ by using the substitutions of $x_i$ variables by the $y_{00}, y_{01}, \ldots, y_{0h}, y_{1h}$ described in Lemma \ref{lem:variable-reduction}. More formally, we define 
\[
\H_{n,h,s} = \{(p_1, \ldots, p_n) : \ubar{q}\in\H' ; p_i = \sum_{j=0}^h q_{j0}q^i_{j1}q_{j0}\}. 
\]\qed

An immediate corollary is the following. 
\begin{corollary}\label{cor:final-derandomization}
The hitting set size and the construction time is 
$s^{(\log s)^{O(1)}}$ for $h=O(1)$. If we want to maintain a subexponential-size hitting set of size $2^{s^{\delta}}$ for $\delta\in (0,1)$, then $h$ can be taken to be at most $c_{\delta}\left(\frac{\log s}{\log\log s}\right)$ where $c_{\delta}$ is a constant depending on $\delta$.   
\end{corollary}

\section{Computing the Matrix Rank over the Free Skew Field}\label{rank-skew-field}
In this section, we give an efficient algorithm to compute the rank of any matrix over the free skew field whose entries are noncommutative polynomials or rational functions with small linear pencils. Additionally we output a matrix tuple on which the rank is achieved. This is done in two steps. Firstly in Section~\ref{sec:blow-up-rational}, we introduce a blow-up definition for matrix rank over the free skew field extending the results for linear pencils. Next, we show an efficient reduction from the rank computation over the free skew field to the linear case in Section~\ref{sec:rank-computation}. The blow-up definition is used to compute the matrix tuple as the witness of the noncommutative rank for such matrices.  

\subsection{On Blow-up Rank of Matrices over the free skew field}\label{sec:blow-up-rational}

We consider a blow-up definition of noncommutative rank for matrices, denoted $\ncrank^*$, over the free skew field. This notion was introduced for linear matrices \cite{DM17, IQS18}, and for any linear matrix $M$ it coincides with $\ncrank(M)$, the inner 
rank of $M$ \cite{IQS18}. In this section we show that $\ncrank^*$ coincides with the inner rank ($\ncrank$) for matrices over the free skew field. We focus on square matrices.


Given a matrix $M=(g_{i,j})_{1\leq i,j\leq m}$ over $\F\newbrak{x_1,\ldots, x_n}$ and $d\in \N$, let 
\[
M^{\{d\}}\ =\ \{M(p_1, \ldots, p_n) \mid
\ (p_1, \ldots, p_n) \in \M^n_d(\F)\}.
\]  
Define $\rank(M^{\{d\}})= \max_{(p_1, \ldots, p_n)} \{\rank(M(p_1, \ldots, p_n))\}.$  We show 
that $\rank(M^{\{d\}})$ is always a multiple of $d$ (Lemma \ref{lem:rat-ex-regularity-lemma}). Moreover, this maximum is achieved for the generic matrix of dimension $d\times d$ as shown in Claim \ref{claim:max-rank-at-generic}.  

\begin{definition}\label{defn:blow-up-rank}
The blow-up rank of the matrix $M$ is defined as 
\[
\ncrank^*(M)=\lim_{d\rightarrow \infty} \frac{\rank(M^{\{d\}})}{d}.
\]

\end{definition}

We first show the existence of this limit and then argue that $\ncrank^*(M)=\ncrank(M)$.

\begin{claim}\label{claim:max-rank-at-generic}
For any $m\times m$ matrix $M = (g_{ij})$ over $\F\newbrak{x_1,\ldots, x_n}$, and for each $d\in \mathbb{N}$, the maximum rank of the image of $M$ for substitutions from $\M^{n}_{d}(\F)$ is the rank of $M$ on $d\times d$ generic matrices. 
\end{claim}

\begin{proof}
Let $\ubar{p}$ in $\M^{n}_{d}(\F)$ be the matrix substitution such that $\rank(M(\ubar{p})) = r$ is maximum. Let 
\[
\rank(M(\ubar{T}))=r^*,
\]
where $\ubar{T} = (T_1, \ldots, T_n)$ is an $n$-tuple of $d\times d$ generic matrices. More precisely, $T_k = \left({t^{(k)}_{i,j}}\right)_{1\leq i,j\leq d}$ and $1\leq k\leq n$.   

Observe that there is a $r\times r$ submatrix of $M(\ubar{p})$ with nonzero determinant. Hence, the determinant of the corresponding submatrix in $M(\ubar{T})$ is a nonzero polynomial over $t^{(k)}_{ij}$ variables. Therefore, $r^*\ge r$.

Conversely, there is an $r^*\times r^*$ submatrix of $M(\ubar{T})$ whose determinant is nonzero. This determinant is a nonzero polynomial in the $t^{(k)}_{i,j}$ variables. Hence, there is a scalar substitution $\ubar{p}$ in $\M^n_d(\F)$ 
for these variables such that the determinant remains nonzero. Clearly, $M(\ubar{p})$ is of rank $r^*$. Therefore, $r\ge r^*$.
\end{proof}

Next, we observe that the regularity lemma~\cite{IQS18}, originally shown for linear matrices, extends to all matrices 
over the free skew field.

\begin{lemma}[A generalization of regularity lemma]\label{lem:rat-ex-regularity-lemma}
For any $m\times m$ matrix $M = (g_{ij})_{1\leq i,j\leq m}$ over $\F\newbrak{x_1,\ldots, x_n}$ there is a positive integer $d_0$ such that the maximum rank of the image of $M$ on $d \times d$ matrix algebra for all $d\geq d_0$ is always a multiple of $d$. 
\end{lemma}

\begin{proof}
The proof is straightforward adaptation of the proof for the linear case as presented in Makam's thesis~\cite[Chapter 4]{Makam18}. 
Let $d_0$ be the minimum positive integer such that any nonzero element $g_{ij}$ is not identically zero on $\M_{d}(\F)$ for $d\geq d_0$.  
By Amitsur's theorem (Theorem \ref{amitsur-domain}) the images of the rational expressions $g_{ij}$ are in the universal division algebra $U(d)$. 
Let $\ubar{T}$ be a tuple of generic matrices of dimension $d\times d$. Then by row and column operations in $U(d)$ it is possible to transform the matrix $M(\ubar{T})$ into the following form: 

\[
 \begin{bmatrix}
I_{d}\\
&I_{d}\\
&&\ddots\\
&&&&0
\end{bmatrix}.
\]
If there are exactly $r$ blocks of $I_{d}$ in the above matrix then clearly its rank is $rd$.  
\end{proof}


\begin{claim}\label{claim:weakly-equivalence}
For any $m\times m$ matrix $M$ over $\F\newbrak{\ubar{x}}$ there is a $d_0\in\mathbb{N}$ such that for $d\ge d_0$ if
$\ubar{T} = (T_1, \ldots, T_n)$ is a tuple of generic matrices of size $d$ and $\ubar{T'} = (T'_1, \ldots, T'_n)$ is a tuple of generic matrices of size $d+1$, then we have that 
\[
\rank(M(T'_1, \ldots, T'_n)) \geq \rank(M(T_1, \ldots, T_n)).
\]
\end{claim}
\begin{proof}
We prove it by induction on $m$. Let $d_0$ be the minimum integer more than $m$ such that any nonzero entry in $M$ is not an identity for the matrix algebra $\M_{d_0}(\F)$. The case $m=1$ follows from Amitsur's theorem on universal division algebra (Theorem \ref{amitsur-domain}) as the image must be invertible.
For the induction, suppose $a_{ij}$ is a nonzero entry of $M$ (it must have a nonzero entry). By row and column permutations
we can rewrite $M$ as:
\[
M = \begin{pmatrix}
a_{ij} &b_i\\
c_j &M'
\end{pmatrix},
\]
where $M'$ is an $(m-1)\times (m-1)$ submatrix of $M$.

Let $M'' = M' - c_ja^{-1}_{ij}b_i \in \M_{m-1}(\F\newbrak{\ubar{x}})$. By some row and column operations we obtain
\[
M = U \begin{pmatrix}
a_{ij} &0\\
0 &M''
\end{pmatrix} V, 
\]
for invertible matrices $U$ and $V$ over $\F\newbrak{\ubar{x}}$. Therefore, for any matrix substitution $\ubar{p}$,
\[
\rank(M(\ubar{p})) = \rank(a_{ij}(\ubar{p})) + \rank(M''(\ubar{p})).
\]
By the induction hypothesis, $\rank(M''(\ubar{T'}))\geq \rank(M''(\ubar{T}))$. Again by Amitsur's theorem, $\rank(a_{ij}(\ubar{T'}))\geq \rank(a_{ij}(\ubar{T}))$. Now the claim follows.
\end{proof}

Notice that, $\rank(M^{\{d\}}) = \rank(M(\ubar{T}))$ and $\rank(M^{\{d+1\}}) = \rank(M(\ubar{T'}))$.
Therefore, \[r_{d+1}(d+1)\geq r_d d > (r_d - 1)(d+1).\] The second inequality follows from the assumption that $d$ is more than $m$. Hence, $r_{d+1} \geq r_d$. The sequence $\{r_d\}$ is then weakly increasing and bounded. Therefore the limit exists.

It follows that for any matrix $M$ over $\F\newbrak{x_1,\ldots,x_m}$ we have
\[
\ncrank^*(M)=\lim_{d\rightarrow \infty} \frac{\rank(M^{\{d\}})}{d}=\max\limits_{d}\frac{\rank(M^{\{d\}})}{d}.
\]

\begin{lemma}\label{lemma-defn-equiv}
For any $m\times m$ matrix $M$ over the free skew field $\F\newbrak{x_1,\ldots,x_n}$ we have
$\ncrank(M)=\ncrank^{*}(M)$. 
\end{lemma}

\begin{proof}
Let $\ncrank(M)=r$. Then there is an $m\times r$ matrix $A$ and $r\times m$ matrix $B$ over $\F\newbrak{x_1,\ldots,x_n}$ such that $M=A\cdot B$. For any matrix substitution  $(p_1, \ldots, p_n)$ of dimension $d\times d$, 
\[
M(p_1, \ldots, p_n)=A(p_1, \ldots, p_n)\cdot B(p_1, \ldots, p_n).\] \[\text{Hence, } \rank(M(p_1, \ldots, p_n))\leq \min\{\rank(A(p_1, \ldots, p_n)),\rank(B(p_1, \ldots, p_n))\}\leq r d.
\]
Therefore, $\ncrank^*(M)\leq r$.

For the other direction, again let $\ncrank(M)= r$. We need to show that for some $d_0\in\mathbb{N}$ and all $d\ge d_0$, for 
the $d\times d$ generic matrix substitution $(T_1,T_2 \ldots,T_n)$ we have $\rank(M(T_1, \ldots, T_n))$ is at least $r d$. 

\begin{claim}\label{clm:rankrelation}
For each $r'\le r$ there is an $r'\times r'$ submatrix $M_{r'}$ of $M$ such that $M_{r'}(T_1, \ldots, T_n)$ is invertible.
\end{claim}

\claimproof
We will prove it by induction on $r'$. For $r'=1$ we can choose any nonzero entry $g_{i,j}$ of the matrix $M$. For sufficiently large 
$d$ the matrix $g_{i,j}(T_1, \ldots, T_n)$ is invertible by Amitsur's theorem (Theorem \ref{amitsur-domain}). 

By induction hypothesis, let $r'=r-1$ and suppose there is an $r'\times r'$ submatrix $M_{r'}$ of $M$ such that 
$M_{r'}(T_1, \ldots, T_n)$ is invertible. Permuting rows and columns suitably we may assume $M_{r'}$ is the top left submatrix indexed by $\{1,2,\ldots,r'\}$. Without loss of generality, we can write:
\[
M =
\left(
\begin{array}{c|c}
M_{r'} & A \\
\hline
B & C
\end{array}
\right). 
\]

Let $C'=C-B{M^{-1}_{r'}}A$ where ${M^{-1}_{r'}}$ is the inverse of $M_{r'}$ over $\F\newbrak{x_1,\ldots,x_n}$.
Then there are invertible matrices $U,V$ corresponding to row and column operations such that
\[
M =
U\left(
\begin{array}{c|c}
M_{r'} & 0 \\
\hline
0 & C'
\end{array}
\right) V. 
\] 

Observe that, $r=\ncrank(M)=\ncrank(M_{r'})+\ncrank(C')=r'+\ncrank(C')$. Hence, $\ncrank(C')>0$. Let $c'_{ij}$ be
a nonzero element of $C'$. Define the matrix $M_r$ of dimension $r\times r$:
\[
M_r =
\left(
\begin{array}{c|c}
M_{r'} & a_j \\
\hline
b_i & c_{i,j}
\end{array}
\right), 
\]
where $b_i$ is $i^{th}$ row of $B$ and $a_j$ is $j^{th}$ column of $A$. As $c'_{ij}= c_{ij}-b_i{M_{r'}}^{-1}a_j\ne 0$ and $d$ is sufficiently large, by Amitsur's
theorem (Theorem \ref{amitsur-domain}) it follows that $c'_{ij}(T_1,T_2,\ldots,T_n)$ is invertible. Hence $M_r(T_1,T_2, \ldots,T_n)$ is invertible which proves the claim.
\qed  

Claim \ref{clm:rankrelation}, shows that there is a submatrix $M_r$ of $M$ such that 
the rank of $M_r(T_1, \ldots, T_n)$ is $rd$. Hence $\ncrank^*(M)\geq r$. 
Putting it together we have shown $\ncrank^*(M)=\ncrank(M)$ which completes the proof.
\end{proof}


\subsection{The Rank Computation}\label{sec:rank-computation}

In this section, we prove Theorem~\ref{thm-nc-rank}. The idea is to reduce the computation of noncommutative rank of a matrix with skew field entries to noncommutative rank computation of a linear matrix incurring a small blow-up in the size. To show the reduction, we need the following lemma.

\begin{lemma}\label{lemma-nc-rank2by2}
Let $P\in \F\newbrak{\ubar{x}}^{m\times m}$ such that,
\[
P = \begin{bmatrix}
A &B\\
C &D
\end{bmatrix},
\]
where $A\in \F\newbrak{\ubar{x}}^{r\times r}$ is invertible. Then,
\[
\ncrank(P) = r+ \ncrank(D - CA^{-1}B),
\]
\end{lemma}

\begin{proof}
If $Q$ is an $n\times n$ invertible matrix over $\F\newbrak{\ubar{x}}$
then
\[
\ncrank(QP)=\ncrank(PQ)=\ncrank(P).
\]
For if $P=MN$ then $QP=(QM)N$ and if $QP=MN$ then $P=(Q^{-1}M)N$.
Similarly for $PQ$.

The matrix
\[
\begin{bmatrix}
A^{-1} &0\\
0 &I_{m-r}
\end{bmatrix}
\]
is full rank. Similarly, the matrix 
\[
\begin{bmatrix}
I_r &0\\
-C &I_{m-r}
\end{bmatrix}
\]
is full rank because
\[
\begin{bmatrix}
I_r &0\\
-C &I_{m-r}
\end{bmatrix}
\begin{bmatrix}
I_r &0\\
C &I_{m-r}
\end{bmatrix} =
\begin{bmatrix}
I_r &0\\
0 &I_{m-r}
\end{bmatrix}.
\]
Hence, $\ncrank(P)$ equals $\ncrank(R)$ where
\[
R= \begin{bmatrix}
I_r &0\\
-C &I_{m-r}
\end{bmatrix}\cdot
\begin{bmatrix}
A^{-1} &0\\
0 &I_{m-r}
\end{bmatrix}\cdot 
\begin{bmatrix}
A &B\\
C &D
\end{bmatrix} =
\begin{bmatrix}
I_r &A^{-1}B\\
0 & D- CA^{-1}B
\end{bmatrix}
\]

Post-multiplying by the invertible matrix
$\begin{bmatrix}
I_r &-A^{-1}B\\
0 &I_{m-r}
\end{bmatrix}$ we obtain
$\begin{bmatrix}
I_r &0\\
0 &D-CA^{-1}B
\end{bmatrix}$.

It is easy to see that its inner rank is $r+\ncrank(D-CA^{-1}B)$.
\end{proof}

In the following lemma, we relate the noncommutative rank of a matrix with skew field entries with small linear pencils to the noncommutative rank of a linear matrix.

\begin{lemma}\label{lemma-ncrank-connection}
Let  $M \in \F\newbrak{\ubar{x}}^{m\times m}$ be a matrix where each $(i,j)^{th}$ entry $g_{ij}$ is computed as the $(1,1)^{th}$ entry of the inverse of a linear pencil $L_{ij}$ of size at most $s$. Then, one can construct a linear pencil $L$ of size $m^2s + m$ such that,
\[
\ncrank(L) = m^2s + \ncrank(M).
\]
\end{lemma}

\begin{proof}
We first describe the construction of the linear pencil $L$ and then argue the correctness. W.l.o.g. we may assume that each linear matrix $L_{ij}$ is $s\times s$ (by padding it, if required, with an identity matrix of suitable size). 

\begin{equation}\label{eq:ncrank-gen}
\text{Let}\quad
L = \left[\begin{array}{ c c c c | c}
L_{11} &0  &\cdots &0  &B_{11}\\
0   &L_{12} &\cdots &0  &B_{12}\\
\vdots   &\vdots &\ddots  &\vdots &\vdots\\
0   &0  &\cdots &L_{mm} &B_{mm}\\
\hline
-C_{11} &-C_{12} &\cdots &-C_{mm} &0
\end{array}
\right], 
\end{equation}
where each $C_{ij}$ is an $m\times s$ and $B_{ij}$ is an $s\times m$ rectangular matrix defined below. Let $e_i$ denote the column vector with 1 in the $i^{th}$ entry and the remaining entries are zero. We define
\[
C_{ij} = \left[\begin{array}{ c| c| c| c}
   & &  &\\
   & &  &\\
   e_i  &0 &\cdots  &0\\
   & &  &\\
   & &  &
\end{array}
\right]
\quad\text{and, }\quad
B_{ij} = \left[\begin{array}{ c c c c c}
   & &e^{T}_j & &\\
   \hline
& &0 & &\\
   \hline
   & &\vdots & &\\
   \hline
& &0 & &
\end{array}
\right].
\]
To argue the correctness of the construction, we write $L$ as a $2\times 2$ block matrix. As each $L_{ij}$ is invertible (otherwise $g_{ij}$ would not be defined), the top-left block entry is invertible. Therefore, we can find two invertible matrices $U,V$ implementing the required row and column operations such that,

\[
L = 
U \left[
\begin{array}{c c c c | c}
L_{11} &0  &\cdots &0  &0\\
0   &L_{12} &\cdots &0  &0\\
\vdots   &\vdots &\ddots  &\vdots &\vdots\\
0   &0  &\cdots &L_{mm} &0\\
\hline
0 &0 &\cdots &0 &\widetilde{D}
\end{array}
\right]V, 
\]
for some $m\times m$ matrix $\widetilde{D}$.

\begin{claim}
The matrix $\widetilde{D}$ is exactly the input matrix $M$.
\end{claim}

\claimproof
From the $2\times 2$ block decomposition we can write,
\[
\widetilde{D} = [C_{11}  C_{12}  \cdots  C_{mm}] 
\left[
\begin{array}{ c c c c}
L^{-1}_{11} &0  &\cdots &0  \\
0   &L^{-1}_{12} &\cdots &0 \\
\vdots   &\vdots &\ddots  &\vdots \\
0   &0  &\cdots &L^{-1}_{mm}
\end{array}
\right]
\left[
\begin{array}{c}
B_{11}\\
B_{12}\\
\vdots\\
B_{mm}
\end{array}
\right]
 = 
 \sum_{i,j} C_{ij}L^{-1}_{ij}B_{ij}.
 \]
Observe that, for each $i,j$, $C_{ij}L^{-1}_{ij}B_{ij}$ is an $m\times m$ matrix with $g_{ij}$ as the $(i,j)^{th}$ entry and remaining entries are 0. Hence, $\widetilde{D} = M$. \qed

Notice that the top-left block of $L$ in Equation~\ref{eq:ncrank-gen} is invertible as for each $i,j\in [m]$, $L_{ij}$ is invertible. Now the proof follows from Lemma~\ref{lemma-nc-rank2by2}.
\end{proof}

\noindent\textbf{Proof of Theorem~\ref{thm-nc-rank}.} For any matrix $M = (g_{i,j})_{m\times m}$ such that for each $i,j\in [m]$, $g_{ij}$ in $\F\newbrak{x_1,\ldots, x_n}$ has a linear pencil of size at most $s$, construct a linear matrix $L$ of size $m^2s+m$ from the previous lemma. We can now compute the noncommutative rank of $L$ using the algorithm of \cite{IQS18} in deterministic $\poly(s,m,n)$-time. Let the rank be $r$. We now output $r-m^2s$ to be the noncommutative rank of $M$. The correctness of the algorithm follows from Lemma~\ref{lemma-ncrank-connection}. 

By the equivalence of the inner rank and blow-up rank established in Section \ref{sec:blow-up-rational}, we know that $\ncrank^*{(M)}=r - m^2s$. Now we use the algorithm in \cite{IQS18} to compute a matrix tuple $\ubar{p}\in \M_d(\F)$ such that the rank of $L(\ubar{p})=rd$ for some $d = O(m^2s)$. Clearly $\rank(M(\ubar{p})) = (r-m^2s)d$. Therefore, the matrix tuple $\ubar{p}$ is also a witness of the rank of $M$. 
\qed


\section{Efficient Linear Pencils for Inversely Disjoint r-Skewed Circuits}\label{expressive-power} 
We now prove that an inversely disjoint rational r-skewed circuit of size $s$ has a linear pencil representation of size $O(s^2)$.
We first prove a  more general result, a composition lemma for linear pencils which implies Theorem \ref{theorem-pencil-power}. 

\begin{lemma}\label{prop-compose}
Let $L$ be an $s\times s$ linear pencil over $x_1, \ldots, x_n$ and
$y_1, \ldots, y_m$. Let $f_{i, j} = (L^{-1})_{i, j}$ for $i,j \in
[s]$. Let $g_1, \ldots, g_m$ be rational
functions over $x_1, \ldots, x_n$ such that each $g_k$ has a linear pencil $L_k$ of size at most
$s_k$. Then we can construct a single linear pencil $\widetilde{L}$ of
size $\sum_{i=1}^m s_i + m + 2s^2+ s$ in $\poly(s_1, \ldots, s_m, s, m, n)$-time such that
\[
(\widetilde{L}^{-1})_{2s^2 + \widehat{s}+i, 2s^2 + \widehat{s}+j} = f_{i,j} (\ubar{x}, g^{-1}_1, \ldots, g^{-1}_m)
\quad\text{ for } i,j\in[s], \text{ where } \widehat{s} = \sum_{i=1}^m s_i + m.
\]
\end{lemma}

\begin{proof}
For each $i,j\in [s]$, $f_{i,j} = (L^{-1}(\ubar{x}, y_1, \ldots,
y_m))_{(i,j)}$ and define $h_{i,j} = f_{i,j}(\ubar{x}, g^{-1}_1, \ldots, g^{-1}_m)$. As the variables $y_1, \ldots, y_m$ are indeterminates, we can
rewrite each $h_{i,j}$ as the following:
\[
h_{i,j} = (L^{-1}(\ubar{x}, g^{-1}_1, \ldots, g^{-1}_m))_{(i,j)}.
\]


We first describe the construction of the linear pencil $\widetilde{L}$
and then prove the correctness of the construction. Let $\widehat{L}$ be a
linear pencil over $x_1, \ldots, x_n$ of size $\widehat{s}$ 
where for each $k\in [m]$, there
exists $i_k,j_k \in [\widehat{s}]$ such that $g^{-1}_k =
(\widehat{L}^{-1})_{i_k, j_k}$. 
The description of $\widehat{L}$ is given later.

Let us first define two $s\times s$ linear
pencils $L'$ and $L''$ as follows. Fix $i,j\in [s]$. Let $(L)_{i, j} =
\alpha_0 + \sum_{k=1}^n \alpha_{k, i, j} x_k + \sum_{k=1}^m \beta_{k,
  i, j} y_k$ . Write $L = L' + L''$ such that $(L')_{i,j}= \alpha_0 +
\sum_{i=k}^n \alpha_{k, i, j} x_k$ and $ (L'')_{i, j} = \sum_{k=1}^m
\beta_{k, i, j} y_k$. We now define $\widetilde{L}$ as a $4\times 4$ block linear matrix of size $\widehat{s}+ 2s^2+ s$,

\begin{equation}\label{eqn-pencil-construction}
\widetilde{L} = 
\left[
\begin{array}{c | c | c | c}
I_{s^2} & A_1 &0 & 0\\
\hline
0 & \widehat{L} &A_2 & 0\\
\hline
0 & 0 & I_{s^2} & A_3\\
\hline
A_4 & 0 & 0 &L'
\end{array}
\right],
\end{equation}

where $I_{s^2}$ is the identity matrix of size $s^2$ and $A_1, A_2, A_3$ and $A_4$ are some rectangular matrices of dimension $s^2\times \hat{s}$, $\hat{s}\times s^2$, $s^2\times s$ and $s\times s^2$ respectively. We now define the construction $A_1, A_2, A_3$ and $A_4$. Subsequently in this proof $I$ is used for $I_{s^2}$. 

\begin{equation*}
\text{Let} \quad \widetilde{L}_1 = 
\left[
\begin{array}{c | c}
I & A_1 \\
\hline
0 & \widehat{L} 
\end{array}
\right]. \quad\quad\text{Then}\quad
\widetilde{L}^{-1}_1 = 
\left[
\begin{array}{c | c}
I & -A_1\widehat{L}^{-1} \\
\hline
0 & \widehat{L}^{-1} 
\end{array}
\right].
\end{equation*}
We now consider the top-left $3\times 3$ block matrix.
\begin{equation*}
\text{Let} \quad \widetilde{L}_2 = 
\left[
\begin{array}{c  c | c}
I & A_1  & 0\\
0 & \widehat{L} & A_2\\
\hline
0 & 0 & I
\end{array}
\right]. \quad\quad\text{Then}\quad
\widetilde{L}^{-1}_2 = 
\left[
\begin{array}{c | c}
\widetilde{L}^{-1}_1 & B_1 \\
\hline
0 & I 
\end{array}
\right],
\end{equation*}
\[
\text{where}\quad B_1 = -
\begin{bmatrix}
I &A_1\\
0 &\widehat{L}
\end{bmatrix}
^{-1}\cdot
\begin{bmatrix}
0\\
A_2
\end{bmatrix}
=
\begin{bmatrix}
A_1\widehat{L}^{-1}A_2\\
-\widehat{L}^{-1}A_2
\end{bmatrix}.
\]
Define the $s^2\times s^2$ matrix $ A_1 \widehat{L}^{-1} A_2 = B_2$. Recall that, $ (L'')_{i, j} = \sum_{k=1}^m \beta_{k, i, j} y_k$. We index the rows of $A_1$ and columns of $A_2$ as a pair $(i,j)$ for some $i,j \in [s]$. Define for each $(i,j)\in [s]\times [s]$ and $k\in [m]$, $(A_1)_{(i,j), i_k} = \beta_{k, i, j}$, $(A_2)_{j_k,(i,j)} = 1$ and the other entries are zero. 
Then, 
\[
(B_2)_{(i,j),(i,j)} = \sum_{i_k,j_k}(A_1)_{(i,j),i_k} (\widehat{L}^{-1})_{i_k,j_k} (A_2)_{j_k,(i,j)} = \sum_{k=1}^m \beta_{k, i, j} g^{-1}_k.
\]
We now define, for each $i,j\in [s]$, $(A_4)_{i,(i,j)} = -1$ and 0 otherwise and $(A_3)_{(i,j),j} = 1$ and 0 otherwise. Since
\begin{equation}\label{eqn-pencil-block}
\widetilde{L} = 
\left[
\begin{array}{c  c  c | c}
I & A_1 &0 & 0\\
0 & \widehat{L} &A_2 & 0\\
0 & 0 & I & A_3\\
\hline
A_4 & 0 & 0 &L'
\end{array}
\right],
\end{equation}
\[
\text{Now,}\quad
\widetilde{L}^{-1} = 
\left[
\begin{array}{c | c}
* & * \\
\hline
* & B_3 
\end{array}
\right]
\quad\quad\quad\text{where,}\quad
B_3 = \left[L' -
\begin{pmatrix}
A_4 & 0 & 0
\end{pmatrix}
\widetilde{L}^{-1}_2
\begin{pmatrix}
0\\
0\\
A_3
\end{pmatrix}\right]^{-1}.
\]
Simplifying further, 
\[
B_3 = (L'-A_4 B_2 A_3)^{-1}  = L^{-1}(\ubar{x}, g^{-1}_1, \ldots, g^{-1}_k).
\]
Therefore, for each $i,j \in [s]$, $(B_3)_{i,j} = (L^{-1}(\ubar{x}, g^{-1}_1, \ldots, g^{-1}_k))_{i,j} = h_{i,j}$. 

Now we construct the linear pencil 
$\widehat{L}$ of size $\hat{s} = \sum_{k=1}^m s_k + m$.

For $k\in [m]$, let there are indices $i'_k,j'_k \in [s_k]$ such that  $g_k = (L^{-1}_k)_{i'_k, j'_k}$. We now define for each $k\in [m]$,

\[
\widetilde{L_k} :=
\left[
\begin{array}{c|c}
L_k & e_{j'_k} \\
\hline
-e^T_{i'_k} & 0
\end{array}
\right]. 
\]
Here the vectors $e_i$ are the unit vector. The construction of $\widehat{L}$ is now as follows:
\begin{equation}\label{eqn-pencil-inverse}
\widehat{L} = 
\begin{bmatrix}
\widetilde{L_1} &0 &\ldots &0\\
0 &\widetilde{L_2} &\ldots &0\\
\vdots &\vdots &\ddots &\vdots\\
0 &0 &\ldots &\widetilde{L_m}
\end{bmatrix}.
\end{equation}

Considering the $\widehat{L}^{-1}$ as an $m\times m$ block matrix where the $i^{th}$ block is of size $s_i+1$, it is easy to see that for each $k\in [m]$, the bottom-right corner entry of the $k^{th}$ block of $\widehat{L}^{-1}$ is $g^{-1}_k$.
To see this apply Equation \ref{2by2inverse} with $p_4=0$. 
\end{proof}

Now the proof of Theorem~\ref{theorem-pencil-power} 
follows easily from Lemma \ref{prop-compose}.
\vspace{0.2cm}

\tkproofpencil.
~~We show that inversely disjoint r-skewed rational functions of height $h$ and of size $s$ have linear pencils of size at most $c s^2$ for some constant $c$. We prove it by induction on the inversion height $h$.
For the base case $h=0$, the input circuit is a noncommutative ABP and the theorem holds by Proposition \ref{prop:abp-pencil}. 

Let $f(\ubar{x}, g_1^{-1}, \ldots, g_m^{-1})$ be an input inversely disjoint r-skewed rational function of height $h$ computed by the circuit $C'$. Replacing $g_i^{-1}$ by new variable $y_i$ we get a noncommutative ABP $C'(\ubar{x}, \ubar{y})$ of size $s'\leq s$. Again by Proposition \ref{prop:abp-pencil}, $C'$ can be represented by a linear pencil of size at most $2s'$. Let $g_1, \ldots, g_m$ are computed by inversely disjoint r-skewed circuits of size $s_1, \ldots, s_m$ and inversion heights $\leq h-1$. By the inductive hypothesis each $g_k$ is computable by a linear pencil of size at most $c s_k^2$.  

Hence by Lemma \ref{prop-compose}, there is a linear pencil of size $S$ representing $C'(\ubar{x}, g_1^{-1}, \ldots, g_m^{-1})$ which satisfies the following condition.
\[
S \leq c \sum_{k=1}^m s_k^2 + m + 8 s'^2 + 2s'.
\]
Simplifying further, 
\[
S\leq c \left(\sum_{k=1}^m s_k^2 + m + s'^2\right), 
\]
for sufficiently large $c$. Since the sub-circuits for $g_1, \ldots, g_m$ are disjoint, we get that  
$(\sum_{k=1}^m s_k^2 + m + s'^2)\leq (\sum_{k=1}^m s_k + m + s')^2\leq s^2$. So, $S\leq c s^2$ for some large constant $c$.
\qed

We now prove the following property of 
the linear pencil constructed in 
Theorem \ref{theorem-pencil-power}. 

\begin{proposition}\label{appendix:property-linear-pencil}
For any inversely disjoint rational r-skewed circuit computing $\r\in \F\newbrak{\ubar{x}}$ and a tuple of matrix $\ubar{p}\in \M^n_m(\F)$ for some finite $m$, the following are equivalent.
\begin{enumerate}
\item $\r$ is defined at $\ubar{p}$.
\item For every gate $u$ which is an output gate or a child of an inverse gate, the pencil constructed in 
Theorem~\ref{theorem-pencil-power} corresponding to 
the rational expression computed at $u$ is invertible at  $\ubar{p}$.
\end{enumerate}
\end{proposition}

\begin{proof}
We first prove that (1)$\implies$(2) by induction on inversion height $h$ of $\r$. For $h=0$, the rational expression $\r$ is a polynomial $f$ computed by a noncommutative r-skewed circuit (ABP). Note that $f$ is defined everywhere and the linear pencil $L$ constructed in Proposition \ref{prop:abp-pencil} is invertible everywhere. 

Let $\r$ is of inversion height $h$. Write $\r = f(\ubar{x}, g^{-1}_1, \ldots, g^{-1}_m)$ and thinking the place holder variables for $g^{-1}_1, \ldots, g^{-1}_m$ as $y_1, \ldots, y_m$ we get $f(\ubar{x}, y_1, \ldots, y_m)$ which is a noncommutative 
ABP over over $\ubar{x}, y_1, \ldots, y_m$.
The rational functions $g^{-1}_1, \ldots, g^{-1}_m$ are of inversion height $\leq h-1$ (some $g_i$ is of inversion height $h-1$ since $\r$ is of inversion height $h$). 

Let $u_1, \ldots, u_m$ be the set of nodes in the circuit of $\r$ such that each $u_k$ is a child of an inverse gate computing $g_k$. 
Let $L$ be the linear pencil corresponding to $f(\ubar{x}, \ubar{y})$ from Proposition \ref{prop:abp-pencil}. 
For some $\ubar{p}$, let $\r(\ubar{p})$ is defined. Therefore, each $g^{-1}_k$ is also invertible at $\ubar{p}$. From the inductive hypothesis, linear pencil $L_k$ (which is constructed by applying Theorem \ref{theorem-pencil-power}) corresponding to $g_k$ is also invertible at $\ubar{p}$. Consider the construction of $\widehat{L}$ from Equation~\ref{eqn-pencil-inverse}. It is easy to see from the construction that $\widehat{L}$ is also invertible at $\ubar{p}$. 

Let $\widetilde{L}$ be the linear pencil representation obtained for $\r$. We now consider the Equation~\ref{eqn-pencil-block} described in Lemma~\ref{prop-compose}. We can conclude that $\widetilde{L}$ is invertible at $\ubar{p}$ if and only if the bottom-right corner block of the inverse, $B_3$ is defined at $\ubar{p}$ i.e. $L(\ubar{x}, g^{-1}_1, \ldots, g^{-1}_m)$ is invertible at $\ubar{p}$. Define $g_i(\ubar{p}) = p'_i$, and $\ubar{q} = (\ubar{p}, p'_1, \ldots, p'_m)$. Clearly, 
$f(\ubar{q}) = \r(\ubar{p})$. 
Since $L$ is a linear pencil for $f$,  it is invertible everywhere. 
In particular, $L(\ubar{x}, g^{-1}_1, \ldots, g^{-1}_m)$ is invertible at $\ubar{p}$ and hence, $\widetilde{L}$ is invertible at $\ubar{p}$.

The other direction follows closely from the proof of \cite[Proposition~7.1]{HW15}. We briefly discuss it for completeness. If $\r$ is not defined at $\ubar{p}$ then there exists a gate computing some rational function $g^{-1}$ in the circuit for $\r$ such that $g(\ubar{p})$ is defined but not invertible. So by the induction hypothesis, the linear pencil $L_g$ (constructed from Theorem \ref{theorem-pencil-power}) representing $g$ (at the entry $(\ell_1, \ell_2)$) is invertible at $\ubar{p}$. Now consider the linear pencil $\widetilde{L}_g$ for 
$g^{-1}$.
From the decomposition, we observe the following. 

\[
\widetilde{L_g} =
\left(
\begin{array}{c|c}
I & 0 \\
\hline
-e^T_{\ell_1} L_g^{-1} & I
\end{array}
\right)
\left(
\begin{array}{c|c}
I & 0 \\
\hline
0 & e^T_{\ell_1} L_g^{-1} e_{\ell_2}
\end{array}
\right)
\left(
\begin{array}{c|c}
L_g & e_{\ell_2} \\
\hline
0 & I
\end{array}
\right).
\]
If $\widetilde{L}_g(\ubar{p})$ is invertible, then $g(\ubar{p})=e^T_{\ell_1} L_g^{-1} e_{\ell_2}$ is also invertible. Hence $\widetilde{L}_g(\ubar{p})$ is not invertible.
\end{proof}

\tkproofrit.
~~Let $\r(\ubar{x}, g_1^{-1}, \ldots, g_m^{-1})$ be the input inversely disjoint r-skewed circuit of size $s$. By Theorem  \ref{theorem-pencil-power}, we construct a linear pencil $\widetilde{L}$ of size $O(s^2)$ for $\r^{-1}$. Now by Proposition \ref{appendix:property-linear-pencil}, $\r^{-1}$ is defined at $\ubar{p}$ if and only if $\widetilde{L}(\ubar{p})$ is invertible. But $\r$ is nonzero if and only if $\r^{-1}$ is defined~\cite{ami66}. So for nonzero testing of $\r$, it is enough to apply the singularity testing   
algorithms in \cite{IQS18} on the linear pencil $\widetilde{L}$ in white-box case. For the black-box case one can use the algorithm in \cite{DM17}. In fact the result in \cite{IQS18} also gives the dimension upper bound of $O(s^2)$ for the tensoring matrices on which $\widetilde{L}$ should be tested for singularity. This also leads to randomized polynomial-time black-box algorithm that simply substitutes the variables randomly from matrices of dimension $O(s^2)$ over sufficiently large fields.    
\qed

\section{Future Directions}\label{sec:open}
Our work raises the following questions for further research: 

\begin{itemize}
\item The most important question is to obtain an \emph{unconditional derandomization} of the black-box RIT problem. 
The current best known result is a quasipolynomial-time black-box RIT algorithm for rational formulas of inversion height at most two~\cite{ACM22}.  

\item Theorem \ref{bw-derand-intro} opens up a new motivation to further study the Conjecture~\ref{BW-conjecture}. In \cite{AJMR17}, it is shown that a nonzero noncommutative polynomial of sparsity $s$ can not be an identity for some $k=O(\log s)$ dimensional matrix algebra. This solves a special case of the conjecture and the proof uses automata theoretic ideas very crucially. Can we improve these techniques to settle the conjecture completely? 

\item The effective use of Higman's trick has found new applications in randomized polynomial-time factorization algorithm for noncommutative formulas~\cite{AJ22}. The proof of Theorem~\ref{thm-nc-rank} does not use Higman's trick. It would be interesting to see whether such ideas can be applied elsewhere.   

\item Can we exactly characterize (up to a polynomial-size equivalence) the expressive power of linear pencil representations for some sub-class of rational circuits?  In this paper, we show that inversely disjoint r-skewed circuits have polynomial-size linear pencils. This gives $\IDRrSC\subseteq \LR$. It would be very interesting to prove that rational r-skewed circuits can be expressed by polynomial-size linear pencils. In other words, prove that $\RrSC=\LR$. 

\end{itemize}

\bibliography{ref2}

\begin{thebibliography}{10}

\bibitem{ami66}
S.A Amitsur.
\newblock Rational identities and applications to algebra and geometry.
\newblock {\em Journal of Algebra}, 3(3):304 -- 359, 1966.

\bibitem{ACM22}
Vikraman Arvind, Abhranil Chatterjee, and Partha Mukhopadhyay.
\newblock Black-box identity testing of noncommutative rational formulas of
  inversion height two in deterministic quasipolynomial-time.
\newblock {\em CoRR}, abs/2202.05693 (to appear in RANDOM 2022), 2022.
\newblock URL: \url{https://arxiv.org/abs/2202.05693}.

\bibitem{AJ22}
Vikraman Arvind and Pushkar~S. Joglekar.
\newblock On efficient noncommutative polynomial factorization via higman
  linearization.
\newblock In Shachar Lovett, editor, {\em 37th Computational Complexity
  Conference, {CCC} 2022, July 20-23, 2022, Philadelphia, PA, {USA}}, volume
  234 of {\em LIPIcs}, pages 12:1--12:22. Schloss Dagstuhl - Leibniz-Zentrum
  f{\"{u}}r Informatik, 2022.
\newblock URL: \url{https://doi.org/10.4230/LIPIcs.CCC.2022.12}.

\bibitem{AJMR17}
Vikraman Arvind, Pushkar~S. Joglekar, Partha Mukhopadhyay, and S.~Raja.
\newblock Randomized polynomial time identity testing for noncommutative
  circuits.
\newblock In {\em Proceedings of the 49th Annual {ACM} {SIGACT} Symposium on
  Theory of Computing, {STOC} 2017, Montreal, QC, Canada, June 19-23, 2017},
  pages 831--841, 2017.

\bibitem{AMS10}
Vikraman Arvind, Partha Mukhopadhyay, and Srikanth Srinivasan.
\newblock New results on noncommutative and commutative polynomial identity
  testing.
\newblock {\em Computational Complexity}, 19(4):521--558, 2010.
\newblock URL: \url{http://dx.doi.org/10.1007/s00037-010-0299-8}.

\bibitem{Ber76}
George~M Bergman.
\newblock Rational relations and rational identities in division rings.
\newblock {\em Journal of Algebra}, 43(1):252 -- 266, 1976.
\newblock URL:
  \url{http://www.sciencedirect.com/science/article/pii/0021869376901599}.

\bibitem{BR11}
J.~Berstel and C.~Reutenauer.
\newblock {\em Noncommutative Rational Series with Applications}.
\newblock Encyclopedia of Mathematics and its Applications. Cambridge
  University Press, 2011.
\newblock URL: \url{https://books.google.co.in/books?id=LL8Nhn72I\_8C}.

\bibitem{BW05}
Andrej Bogdanov and Hoeteck Wee.
\newblock More on noncommutative polynomial identity testing.
\newblock In {\em 20th Annual {IEEE} Conference on Computational Complexity
  {(CCC} 2005), 11-15 June 2005, San Jose, CA, {USA}}, pages 92--99, 2005.

\bibitem{Chatterjee21}
Prerona Chatterjee.
\newblock Separating abps and some structured formulas in the non-commutative
  setting.
\newblock In Valentine Kabanets, editor, {\em 36th Computational Complexity
  Conference, {CCC} 2021, July 20-23, 2021, Toronto, Ontario, Canada (Virtual
  Conference)}, volume 200 of {\em LIPIcs}, pages 7:1--7:24. Schloss Dagstuhl -
  Leibniz-Zentrum f{\"{u}}r Informatik, 2021.

\bibitem{CKS18}
Chi{-}Ning Chou, Mrinal Kumar, and Noam Solomon.
\newblock Hardness vs randomness for bounded depth arithmetic circuits.
\newblock In Rocco~A. Servedio, editor, {\em 33rd Computational Complexity
  Conference, {CCC} 2018, June 22-24, 2018, San Diego, CA, {USA}}, volume 102
  of {\em LIPIcs}, pages 13:1--13:17. Schloss Dagstuhl - Leibniz-Zentrum
  f{\"{u}}r Informatik, 2018.
\newblock URL: \url{https://doi.org/10.4230/LIPIcs.CCC.2018.13}.

\bibitem{Cohn71}
P.~M. Cohn.
\newblock The embedding of firs in skew fields.
\newblock {\em Proceedings of The London Mathematical Society}, pages 193--213,
  1971.

\bibitem{Cohn95}
P.~M. Cohn.
\newblock {\em Skew Fields: Theory of General Division Rings}.
\newblock Encyclopedia of Mathematics and its Applications. Cambridge
  University Press, 1995.

\bibitem{DM17}
Harm Derksen and Visu Makam.
\newblock Polynomial degree bounds for matrix semi-invariants.
\newblock {\em Advances in Mathematics}, 310:44--63, 2017.

\bibitem{DSY09}
Zeev Dvir, Amir Shpilka, and Amir Yehudayoff.
\newblock Hardness-randomness tradeoffs for bounded depth arithmetic circuits.
\newblock {\em {SIAM} J. Comput.}, 39(4):1279--1293, 2009.
\newblock URL: \url{https://doi.org/10.1137/080735850}.

\bibitem{Eilenberg74}
Samuel Eilenberg.
\newblock {\em Automata, Languages, and Machines (Vol A)}.
\newblock Pure and Applied Mathematics. Academic Press, 1974.

\bibitem{FGT21}
Stephen~A. Fenner, Rohit Gurjar, and Thomas Thierauf.
\newblock Bipartite perfect matching is in quasi-nc.
\newblock {\em {SIAM} J. Comput.}, 50(3), 2021.
\newblock URL: \url{https://doi.org/10.1137/16M1097870}.

\bibitem{FS13}
Michael~A. Forbes and Amir Shpilka.
\newblock Quasipolynomial-time identity testing of non-commutative and
  read-once oblivious algebraic branching programs.
\newblock In {\em 54th Annual {IEEE} Symposium on Foundations of Computer
  Science, {FOCS} 2013, 26-29 October, 2013, Berkeley, CA, {USA}}, pages
  243--252, 2013.

\bibitem{GGOW20}
Ankit Garg, Leonid Gurvits, Rafael~Mendes de~Oliveira, and Avi Wigderson.
\newblock Operator scaling: Theory and applications.
\newblock {\em Found. Comput. Math.}, 20(2):223--290, 2020.
\newblock URL: \url{https://doi.org/10.1007/s10208-019-09417-z}.

\bibitem{GGOW16}
Ankit Garg, Leonid Gurvits, Rafael~Mendes{ de} Oliveira, and Avi Wigderson.
\newblock A deterministic polynomial time algorithm for non-commutative
  rational identity testing.
\newblock {\em 2016 IEEE 57th Annual Symposium on Foundations of Computer
  Science (FOCS)}, pages 109--117, 2016.

\bibitem{HS80}
Joos Heintz and Claus{-}Peter Schnorr.
\newblock Testing polynomials which are easy to compute (extended abstract).
\newblock In Raymond~E. Miller, Seymour Ginsburg, Walter~A. Burkhard, and
  Richard~J. Lipton, editors, {\em Proceedings of the 12th Annual {ACM}
  Symposium on Theory of Computing, April 28-30, 1980, Los Angeles, California,
  {USA}}, pages 262--272. {ACM}, 1980.
\newblock URL: \url{https://doi.org/10.1145/800141.804674}.

\bibitem{Hig40}
Graham Higman.
\newblock {\em Units in group rings}.
\newblock PhD Thesis. 1940.

\bibitem{HW15}
Pavel Hrube{\v{s}} and Avi Wigderson.
\newblock Non-commutative arithmetic circuits with division.
\newblock {\em Theory of Computing}, 11(14):357--393, 2015.
\newblock URL: \url{http://www.theoryofcomputing.org/articles/v011a014}.

\bibitem{Hua49}
Loo-Keng Hua.
\newblock Some properties of a sfield.
\newblock {\em Proceedings of the National Academy of Sciences of the United
  States of America}, 35(9):533--537, 1949.
\newblock URL: \url{http://www.jstor.org/stable/88328}.

\bibitem{IQS18}
G{\'a}bor Ivanyos, Youming Qiao, and K.~V. Subrahmanyam.
\newblock Constructive non-commutative rank computation is in deterministic
  polynomial time.
\newblock {\em computational complexity}, 27(4):561--593, Dec 2018.

\bibitem{KI04}
Valentine Kabanets and Russell Impagliazzo.
\newblock Derandomizing polynomial identity tests means proving circuit lower
  bounds.
\newblock {\em Comput. Complex.}, 13(1-2):1--46, 2004.

\bibitem{KV09}
Dmitry~S. Kaliuzhnyi-Verbovetskyi and Victor Vinnikov.
\newblock Singularities of rational functions and minimal factorizations: The
  noncommutative and the commutative setting.
\newblock {\em Linear Algebra and its Applications}, 430(4):869--889, 2009.
\newblock URL:
  \url{https://www.sciencedirect.com/science/article/pii/S0024379508003893}.

\bibitem{KS01}
Adam~R. Klivans and Daniel Spielman.
\newblock Randomness efficient identity testing of multivariate polynomials.
\newblock In {\em Proceedings of the Thirty-third Annual ACM Symposium on
  Theory of Computing}, STOC '01, pages 216--223, New York, NY, USA, 2001. ACM.

\bibitem{LST21}
Nutan Limaye, Srikanth Srinivasan, and S{\'{e}}bastien Tavenas.
\newblock Superpolynomial lower bounds against low-depth algebraic circuits.
\newblock In {\em 62nd {IEEE} Annual Symposium on Foundations of Computer
  Science, {FOCS} 2021, Denver, CO, USA, February 7-10, 2022}, pages 804--814.
  {IEEE}, 2021.
\newblock URL: \url{https://doi.org/10.1109/FOCS52979.2021.00083}.

\bibitem{Makam18}
Visu Makam.
\newblock {\em Invariant Theory, Tensors and Computational Complexity}.
\newblock PhD Thesis. 2018.

\bibitem{Ni91}
Noam Nisan.
\newblock Lower bounds for non-commutative computation (extended abstract).
\newblock In {\em Proceedings of the 23rd Annual {ACM} Symposium on Theory of
  Computing, May 5-8, 1991, New Orleans, Louisiana, {USA}}, pages 410--418,
  1991.

\bibitem{row80}
Louis~Halle Rowen.
\newblock {\em Polynomial identities in ring theory}.
\newblock Pure and Applied Mathematics. Academic Press, 1980.
\newblock URL:
  \url{http://gen.lib.rus.ec/book/index.php?md5=bde982110d09e6199643e04da0558459}.

\bibitem{Str73}
Volker Strassen.
\newblock Vermeidung von divisionen.
\newblock {\em Journal für die reine und angewandte Mathematik}, 264:184--202,
  1973.
\newblock URL: \url{http://eudml.org/doc/151394}.

\bibitem{LST22}
S{\'{e}}bastien Tavenas, Nutan Limaye, and Srikanth Srinivasan.
\newblock Set-multilinear and non-commutative formula lower bounds for iterated
  matrix multiplication.
\newblock In Stefano Leonardi and Anupam Gupta, editors, {\em {STOC} '22: 54th
  Annual {ACM} {SIGACT} Symposium on Theory of Computing, Rome, Italy, June 20
  - 24, 2022}, pages 416--425. {ACM}, 2022.
\newblock URL: \url{https://doi.org/10.1145/3519935.3520044}.

\bibitem{vol18}
Jurij Vol\v{c}i\v{c}.
\newblock Matrix coefficient realization theory of noncommutative rational
  functions.
\newblock {\em Journal of Algebra}, 499:397--437, 04 2018.

\end{thebibliography}

\end{document}